\documentclass[10pt,journal]{IEEEtran}
\makeatletter
\adddialect\l@ENGLISH\l@english
\makeatother

\usepackage{algorithm}
\usepackage{algpseudocode}
\usepackage{cite}
\usepackage{graphicx}
\usepackage{multirow}
\usepackage{url}
\usepackage{amsmath}
\usepackage{amssymb}
\usepackage{epstopdf}
\usepackage{textcomp}
\usepackage{amsthm}
\usepackage{color}
\usepackage[most]{tcolorbox}
\usepackage{enumitem}
\usepackage{colortbl}

\fontfamily{pcr}\selectfont

\tcbset{myformula/.style={colback=white, %yellow!10!white,
		colframe=black, %red!50!black,
		top=0pt,bottom=0pt,left=0pt,right=2pt,
		boxsep=0pt,
		arc=0pt,
		outer arc=0pt,
}}

\ifCLASSINFOpdf
\else
\fi
\begin{document}

% ------------------------------------Title --------------------------------------------------

\title{SmartCon: Deep Probabilistic Learning Based Intelligent Link-Configuration in Narrowband-IoT Towards 5G and B5G}

%-----------------------------------Author ---------------------------------------------------

\author{Raja~Karmakar,~\IEEEmembership{Member,~IEEE},
     Georges~Kaddoum,~\IEEEmembership{Senior~Member,~IEEE}
     and~Samiran~Chattopadhyay,~\IEEEmembership{Senior~Member,~IEEE}

%\IEEEcompsocitemizethanks{\IEEEcompsocthanksitem M. Shell was with the %Department
%of Electrical and Computer Engineering, Georgia Institute of Technology, Atlanta,
%GA, 30332.\protect\\
 %note need leading \protect in front of \\ to get a newline within \thanks as
 %\\ is fragile and will error, could use \hfil\break instead.
%E-mail: see http://www.michaelshell.org/contact.html
%\IEEEcompsocthanksitem J. Doe and J. Doe are with Anonymous University.}% <-this % stops an unwanted space

\thanks{R. Karmakar and G. Kaddoum are with Department of Electrical Engineering, ETS, University of Quebec, Montreal, Canada (Email: raja.karmakar.1@ens.etsmtl.ca, georges.kaddoum@etsmtl.ca).}
\thanks{S. Chattopadhyay is with Department of Information Technology, Jadavpur University, Kolkata, India 700098, and Institute for Advancing Intelligence, TCG Centres for Research and Education in Science and Technology, Kolkata, India (Email: samiran.chattopadhyay@jadavpuruniversity.in).}}

\maketitle

%\markboth{IEEE Transactions on Cognitive Communications and Networking}{Karmakar \MakeLowercase{\textit{et al.}}}

%\IEEEtitleabstractindextext{%
\begin{abstract}

%Narrowband Internet of Things (NB-IoT) is a new radio access network protocol defined by the Third Generation Partnership Project (3GPP) to provide enhanced coverage and low power consumption. 
To enhance the coverage and transmission reliability, {\em repetitions} adopted by Narrowband Internet of Things (NB-IoT) allow repeating transmissions several times. %, making it cost-effective. 
%To enhance the coverage and transmission reliability, {\em repetitions} adopted by NB-IoT allow repeating transmissions several times.
%In the era of narrow-band radio technology, 
%Narrowband Internet of Things (NB-IoT) is a new radio access network protocol defined by the Third Generation Partnership Project (3GPP) to provide enhanced coverage and low power consumption. %, making it cost-effective. 
%To enhance the coverage and transmission reliability, {\em repetitions} adopted by NB-IoT allow repeating transmissions several times.
%However, when the signal strength is high, the application of repetitions imposes unnecessary transmission that results in waste of radio resources. 
However, this results in a waste of radio resources when the signal strength is high.
In addition, in low signal quality, the selection of a higher {\em modulation and coding scheme (MCS)} level leads to a huge packet loss in the network. %Meanwhile, when the signal strength is high, the application of a lower MCS fails to increase the network performance. %since a low MCS provides a low data rate. %Therefore, the MCS value needs to be chosen dynamically based on the signal quality of the channel.
%and consequently this issue significantly affects the performance of NB-IoT which has . 
Moreover, the number of {\em physical resource blocks (PRBs)} per-user needs to be chosen dynamically, such that the utilization of radio resources can be improved on per-user basis. Therefore, in NB-IoT systems, dynamic adaptation of repetitions, MCS, and radio resources, known as {\em auto link-configuration}, is crucial. %To automate the selection of the aforesaid parameters, we can introduce intelligence to the system to enable it to smartly cope with different network scenarios.
%that needs to be addressed intelligently to cope with different network scenarios. 
Accordingly, in this paper, we propose {\em SmartCon} which is a {\em Generative Adversarial Network (GAN)}-based deep learning approach for auto link-configuration during uplink or downlink scheduling, such that the packet loss rate is significantly reduced in NB-IoT networks.
%Accordingly, in this paper, we propose {\em SmartCon} which is a {\em Generative Adversarial Network (GAN)}-based auto link-configuration mechanism for NB-IoT. SmartCon uses a deep learning approach to dynamically generate the best suited values of MCS, PRB, and repetitions, during uplink or downlink scheduling, such that the packet loss rate is significantly reduced in NB-IoT networks. %In order to find out these values, the GAN model estimates the channel condition based on signal-to-interference-plus-noise ratio (SINR). 
For the training purpose of the GAN, we use a {\em Multi-Armed Bandit (MAB)}-based reinforcement learning mechanism that intelligently tunes its output depending on the present network condition. %By using the MAB, we intelligently select MCS values, repetitions and the number of PRB, during scheduling of packets.
%The proposed mechanism can lead to a smart unique NB-IoT based framework for developing Industrial Internet of Things (IIoT) frameworks in 5G and beyond 5G (B5G). 
The performance of SmartCon is thoroughly evaluated through simulations where it is shown to significantly improve the performance of NB-IoT systems compared to baseline schemes.

\end{abstract}

%\maketitle

%\IEEEdisplaynontitleabstractindextext

% \fi
\IEEEpeerreviewmaketitle

\section{Introduction}
\label{sec:intro}
%The concept of the Internet of Things (IoT)~\cite{bakshi2016emit}, which implies connectivity to a network for everything, has become an important part of our lives. 
The number of Internet of Things (IoT)~\cite{bakshi2016emit} devices is constantly increasing in the fifth-generation (5G) and beyond 5G (B5G) of mobile telecommunications. %IoT devices enable various on-demand application services such as smart homes, environmental monitoring, body/health monitoring, condition-based maintenance, etc. 
To meet the demands described by the IoT specifications, the Third Generation Partnership Project (3GPP) has presented a new radio access technology, known as Narrowband Internet of Things (NB-IoT)~\cite{popli2019survey,3GPPIoT}. %, which leads to a promising footstep towards the evolution of fifth-generation (5G) IoT~\cite{3GPPIoT}. 
NB-IoT can provide an improved coverage compared to Long-Term Evolution (LTE) networks, massive device connectivity, ultra-low device complexity or costs, and low power consumption~\cite{wang2017primer}. Specifically, NB-IoT is a variant of LTE, designed for IoT frameworks. 
%In NB-IoT networks, the allocation of radio resources is more complicated than in LTE.
Like LTE, the NB-IoT technology is based on orthogonal frequency-division multiple access (OFDMA), with a system bandwidth of $180$ kHz which is equal to one {\em physical resource block (PRB)} in 4G LTE transmissions. Given this low channel bandwidth, NB-IoT specifically focuses on indoor coverage, and data transmission with a higher latency~\cite{popli2019survey,3GPPIoT}. %maintain a latency tolerance level of approximately $10$ seconds~\cite{3GPPIoT}. 
%Therefore, NB-IoT primarily focuses on IoT devices that can tolerate transmission delays and are suitable for small range communications. %NB-IoT uses single carrier frequency-division multiple access (SC-FDMA) in the uplink and orthogonal frequency-division multiplexing (OFDM) in the downlink. In addition, more features are included in NB-IoT to fulfill the requirements of IoT-based applications. 
%For downlink transmission, NB-IoT devices primarily perform data transmission on one dimension (time-domain), whereas LTE devices can perform the transmission on both the frequency and time domains (i.e., two dimensions). 
Dynamic adjustment to different radio conditions can be performed by configuring the {\em modulation and coding scheme (MCS)} value, which is defined as the combination of a type of modulation and coding rate used for a given PRB~\cite{rico2016overview,zayas20173gpp}. The MCS is a key feature which is used to set the data rate of a transmission in a wireless connection~\cite{zayas20173gpp}. 
In NB-IoT, the MCS value is between $0$ and $12$, with a variable Transport Block Size (TBS)~\cite{3gppNB-IoT,rico2016overview}. %A higher MCS level reduces the coding redundancy, and thus increases the TBS for the same number of resources.
 %The MCS defines two aspects -- code rate and modulation of a transmission where the number of useful bits can be transfered per {\em resource element (RE)}. 
The MCS also specifies how many bits can be transferred per {\em resource element (RE)} which is the smallest modulation structure in LTE~\cite{3GPPIoT}.

\begin{figure}[!t]
 \centering
 \includegraphics[width=\linewidth]{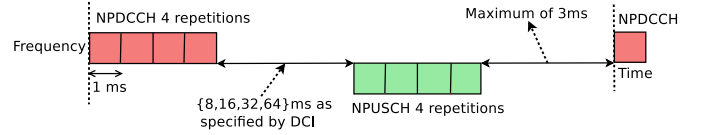}
 \caption{An illustration of repetition}
 \label{repetition}
 \vspace{-5mm}
\end{figure}

In order to achieve coverage enhancement and improve transmit reliability in NB-IoT, the concept of {\em repetitions} is used in the data and control signal transmissions~\cite{ravi2019evaluation,chafii2018enhancing}. Repetitions imply repeating the transmission several times~\cite{zayas20173gpp}. 
%The repetition of data or control signals is considered as a promising mechanism towards the enhancement of coverage and transmit reliability in NB-IoT systems~\cite{ravi2019evaluation,chafii2018enhancing}. %This is because the transmission reliability is enhanced by the use of repetitions, as reported by 3GPP Release 13, and thus repetitions should be enabled when the signal strength is poor~\cite{yu2020npdcch}.
%\begin{figure}[!t]
% \centering
% \includegraphics[width=\linewidth]{figure/repetition.eps}
% \caption{An illustration of repetition}
% \label{repetition}
% %\vspace{-5mm}
%\end{figure}
%In NB-IoT, the repetition is the primary solution in order to provide enhanced coverage with a low complexity~\cite{ravi2019evaluation}. 
%The repetition of a complete transmission can be applied to both control signaling and data transmissions. 
%Before transmitting data, the control information containing the number of resources, selected MCS level and repetition number is transmitted through the proper control channel -- Narrowband Physical Uplink Shared Channel (NPUSCH) (in uplink scheduling) or Narrowband Physical Downlink Control Channel (NPDCCH) (in downlink scheduling). %NPUSCH carries both the control information and data for uplink scheduling, whereas NPDCCH primarily carries the data for downlink scheduling. 
The repetition for the uplink and downlink transmissions can be selected from $\{1,2,4,8,16,32,64,128\}$ and $\{1,2,4,8,16,32,64,128,256,512,1024,2048\}$, respectively, where the selected number denotes the number of repetition of the same transmission block~\cite{zayas20173gpp}. Fig.~\ref{repetition} illustrates a repetition of $4$ in NB-IoT with both Narrowband Physical Uplink Shared Channel (NPUSCH) and Narrowband Physical Downlink Control Channel (NPDCCH) transmission blocks, where the content of each of these blocks is repeated $4$ times during a single transmission. The time gap between the NPDCCH and NPUSCH repetitions is defined by the downlink control information (DCI). It specifies a scheduling index that permits a device to collect data during downlink scheduling~\cite{malik2019radio}.

%we define dynamic {\em link-configuration} as an auto-adaptation of MCS levels, repetition numbers and radio resources during scheduling, and this adaptation is a crucial problem that needs to be addressed precisely in NB-IoT networks. 
%the dynamic adaptation of the repetition number and MCS levels during scheduling is referred to as {\em auto link-configuration} which is a crucial problem to be addressed precisely in NB-IoT. 
%We define dynamic adaptation of repetitions, MCS, and PRB as {\em auto link-configuration} in NB-IoT.
%Considering MCS, the new coverage enhancing feature, and PRB, we can represent the link configuration in NB-IoT as a three dimensional problem -- (i) selection of MCS values, (ii) determination of the repetitions, and (iii) selection of the number of PRB. 
%The dynamic adjustment to different radio conditions can be performed by configuring 

%As reported by the 3GPP Release 13, 
Since the transmission reliability is enhanced by the use of repetitions, it should be enabled when the signal strength is poor~\cite{yu2020npdcch}.
On the other hand, the MCS level needs to be choosen dynamically based on the signal strength~\cite{yu2017uplink}. When the channel conditions are poor, the selection of a high MCS value results in a higher {\em packet loss rate (PLR)}, and consequently the system throughput is reduced. %Conversely, in case of high signal quality, the application of a low MCS improves the transmit reliability and coverage; however, it does not achieve the high data rate. %that can be achieved when the channel condition is good. 
%Therefore, the MCS is dependent on the quality of the radio link where more data can be sent as the channel quality improves.  wirges2019performance
Moreover, the rapid changes in channel conditions lead to high fluctuations in the PLR in NB-IoT networks~\cite{yu2017uplink,wirges2019performance}. Therefore, during scheduling, both MCS and repetitions play a crucial role in the packet transmission such that the best suited data rate, coverage and a low PLR can be achieved based on the present channel condition. Moreover, NB-IoT systems use radio resource blocks reserved by LTE systems~\cite{malik2018radio}, and thus appropriate utilization of radio resources is especially demanded for NB-IoT. Therefore, an adaptive selection of the number of PRBs per-user is also required in NB-IoT. %along with the adjustment of MCS levels and repetitions.

%Considering the aforementioned discussion, dynamic adjustment of these parameters in parallel becomes a challenging issue, especially in NB-IoT systems since given their low bandwidth, less compex hardware, high packet transmission delay, and high packet loss rate in poor signal strength. 
Therefore, during uplink or downlink scheduling, a dynamic adaptation, known as {\em auto link-configuration}, is required for MCS levels, repetitions, and per-user PRB in NB-IoT systems. % and we define this problem as {\em auto link-configuration}.
Therefore, we can represent the auto link-configuration in NB-IoT as a three dimensional problem -- (i) selection of MCS values, (ii) determination of the repetitions, and (iii) selection of the number of PRBs per-user. %As a consequence, in auto link-configuration, these parameters need to be dynamically adjusted in parallel.  

%Due to the simultaneous dynamic adjustment of these parameters, the auto link-configuration becomes challenging and crucial problem that needs to be addressed precisely in NB-IoT networks. 

\subsection{State-of-the-Art}

%The existing works primarily address the scheduling, resource allocation, and energy efficiency in NB-IoT.

%The work~\cite{popli2019survey} presents a survey that summarizes the resource allocation, technical features, energy efficiency, and applications of NB-IoT. In~\cite{feltrin2019narrowband}, the primary features and mechanism of data transmissions over NB-IoT are presented along with the performance analysis of downlink and uplink transmissions. Arbitrary scenarios are considered in~\cite{tong2019positioning} in order to highlight the performance issues that arise due to the variation of positioning of NB-IoT devices. This work also discusses different network factors, such as coverage, locations of devices, etc. Authors in~\cite{da2014internet,sisinni2018industrial} discuss the challenges along with the research directions in Industrial IoT, related to real time performance, energy efficiency, privacy and security, and interoperability and coexistence with existing IoT standards.

%By considering group communications, authors in~\cite{tsoukaneri2018group} propose an extension in the framework of NB-IoT, which incorporates a channel for performing group communications in the presence of interference. 
An effective approach for small data transmission in NB-IoT is proposed in~\cite{oh2016efficient}, without the consideration of the connection setup process related to radio resource control. However, this work does not focus on the adaptation of MCS and repetition number. 
Authors in~\cite{chen2019performance} model the random access traffic in NB-IoT by considering the arrival of processes and their services, where the network delay is analyzed based on random latency bounds.  
%To describe the length of queue and re-transmissions due to packet collisions, Markov chain is exploited in~\cite{sun2017throughput}. This model does not consider the resource allocation with an adaptive MCS. 
The work~\cite{mostafa2017connectivity} discusses the primary challenges of providing a stable connectivity to a huge number of machine-type communication (MTC) devices in NB-IoT networks. 
%Several existing works, such as~\cite{hsieh2018design,ratasuk2016data,yu2017uplink}, consider the design of uplink or downlink scheduling in NB-IoT.
In~\cite{hsieh2018design}, the proposed uplink scheduler for NB-IoT frameworks is a basic threshold-based approach with user equipment (UE) specific requirements, which is mainly suitable for homogeneous traffic. Details and the uplink and downlink transmission channels' performance are discussed in~\cite{ratasuk2016data} with a focus on the design approaches in NB-IoT. Yu \textit{et al.}~\cite{yu2017uplink} propose an uplink scheduling mechanism for NB-IoT, where the uplink link adaptation, including the determination of the MCS value and repetition number, is performed based on the present channel condition. %The transmission gap and power control related uplink scheduling technologies are also analyzed in~\cite{yu2017uplink}. 
However, this work does not consider downlink scheduling and it uses a threshold-based mechanism for the MCS and repetitions selection in NB-IoT systems. 

In the direction of resource management, the work~\cite{malik2018radio} designs a mechanism for resource allocation in NB-IoT, by focusing only on the rate maximization. Manne \textit{et al.}~\cite{manne2020scheduling} explain NPDCCH physical layer procedures with the technique of search space decoding, where a resource mapping scheme is discussed for NPDCCH by utilizing uplink reference signals. %However, only the allotment of search space in NPDCCH is considered by the proposed schedulers. 
%Boisguene \textit{et al.}~\cite{RubbensSurveyNBIOT:2017} consider the design and resource allocation issues in downlink NB-IoT systems. 
The heuristic algorithm proposed in~\cite{yu2018downlink} discusses a downlink scheduling mechanism in NB-IoT. In this work, the objective is to efficiently use radio resources in order to support massive connections in the network.
In the scheme discussed in~\cite{huang2019radio}, narrowband physical downlink shared channel (NPDSCH) subframes are assigned continuously in the radio resource scheduling until a device gets a maximum number of subframes, such that the allocated resources can satisfy the data transmission requirement. %This work specifically describes the issues related to radio resource scheduling over NB-IoT architectures. 
%The work~\cite{yu2020npdcch} proposes a downlink scheduling mechanism in NB-IoT, with a variable NPDSCH period. 
The work~\cite{yu2020npdcch} deals with the enhancement of radio resource utilization for NB-IoT by minimizing the consumption of radio resources during downlink transmission. However, the dynamic adaptation of MCS and repetition number are not addressed in this work.

Considering the power efficiency of NB-IoT systems, the authors in~\cite{elgarhy2020rate} discuss resource allocation during uplink transmission and analyze the trade-off between power, latency, and rate.
The work~\cite{lei2019joint} specifically studies the radio resource allocation with scheduling and computation offloading by focusing on the minimization of the power consumption and average delay in NB-IoT based systems.
%For low power scenarios, the impact of the channel coherence time on the uplink coverage is highlighted in~\cite{beyene2017performance} by addressing repetitions of signal transmissions provided by the end-devices. %In~\cite{azari2018latency}, only the latency-energy tradeoff is presented through a queuing model. 
Although a scheduling is discussed in~\cite{azari2019latency} by considering different coverage classes, latency, and power consumption in NB-IoT, the proposed mechanism does not dynamically adapt the MCS and repetition number during the scheduling. Considering the network slicing in 5G communications, the work~\cite{hua2019gan} addresses the issue of dynamic allocation of resources for different services over a common physical infrastructure. Accordingly, the authors in~\cite{hua2019gan} propose a demand-aware approach for resource allocation in network slicing by combining deep distributional reinforcement learning and GAN.

Therefore, the existing works do not deal with the challenge of intelligent selection of MCS values, repetition numbers, and resources in both uplink and downlink scheduling in NB-IoT. Moreover, the aforesaid parameters have trade-offs, namely when the signal strength is low, the MCS level and number of PRBs need to be decreased but repetitions should be increased. In addition, the selection of these parameters should dynamically cope with different network conditions, considering the constraints (bandwith, delay, PLR, etc.) of NB-IoT devices and without any prior knowledge of the wireless environment. %Thus, the system should apply an online learning approach to learn the environment and accordingly, automatically act by considering the trade-offs of the parameters. 
Consequently, by considering the trade-offs in the MCS, repetitions, and PRB, an online learning based smart technique is required to learn the environment and accordingly, automatically adapt these parameters in parallel. %Therefore, an intelligent mechanism can improve the adaptation of these parameters, leading to the enhancement of the performance of NB-IoT systems.
%such that the packet loss rate will be minimized. 

\subsection{Our Approach}
%by exploiting an online learning based approach, 
In this paper, we propose {\em SmartCon} which is an intelligent adaptation of MCS, repetitions, and PRB during uplink or downlink scheduling, such that the packet loss rate is significantly reduced in NB-IoT networks.
%In this direction, we design a {\em Generative Adversarial Network (GAN)}~\cite{lei2019gcn} to intelligently generate the real dynamics of scheduling events (uplink and downlink) with the best possible MCS levels, number of repetitions and PRB. In order to estimation these values, the GAN model imposes the variation of channel condition as the latent factor, and thus SmartCon is able to capture the inherent dynamics of wireless environments.
In this direction, we design a {\em Generative Adversarial Network (GAN)}~\cite{lei2019gcn} model that uses a deep learning approach to dynamically generate the best suited values for the aforesaid parameters for future scheduling. %, so that the packet loss rate in the network will be significantly reduced. In order to estimate these values 
The proposed GAN considers the variation of signal strength and noise of the channel inputs. %Consequently, SmartCon is able to capture the inherent dynamic nature of wireless environments. %which is due to the signal interference and fluctuation of signal strength in the network. 
%estimates the channel condition based on signal-to-interference-plus-noise ratio (SINR). 
%In addition, we apply {\em temporal point process (TPP)} which is a stochastic process containing isolated events. 
In SmartCon, the {\em temporal point process (TPP)} specifies the sequence of time instances of future scheduling (uplink or downlink) associated with the best possible MCS levels, the number of repetitions, and PRBs. To train the GAN, we use a {\em Multi-Armed Bandit (MAB)} based reinforcement learning mechanism that dynamically tunes its output depending on the impact of the environment. %During packet scheduling, the MAB intelligently selects the MCS, repetitions, and the number of PRBs. 
%The MAB also captures the present signal-to-interference-plus-noise ratio (SINR) of the channel. %, such that the adaptation of MCS, repetitions, and resources can be learnt according to the current channel condition. %The values of the parameters selected under a given SINR are stored along with the SINR value and this entire information serves as the training dataset to train the proposed GAN model in SmartCon. 
To the best of our knowledge, SmartCon is the first work that considers intelligently adapting MCS and repetitions, along with radio resources in NB-IoT systems. %, such that the packet loss rate in the network is reduced.   

\textbf{Reason for applying MAB for generating the training dataset:} The MAB is a reinforcement learning mechanism, where a learning agent opts for a single option (known as {\em arm}) from a set of available options which have unknown characteristics at the initial stage. Based on its choice, a certain {\em reward} is received by the agent. The agent always tries to maximize the cumulative reward. In order to generate the training dataset in our proposed mechanism, each combination of the MCS, repetitions, and PRBs can be considered as an {\em arm} in the MAB. Therefore, to dynamically select values of the MCS, repetitions, and PRBs, the agent needs to select an {\em arm} based on the present channel condition such that the packet loss rate will be minimized. Thus, at any time instance, the arm and the associated channel condition can be considered as a state. Therefore, the MAB is a suitable learning model to populate a dataset containing the information related to the intelligent selection of the MCS, repetitions, and PRBs, considering the signal strength of the channel. Consequently, the generated dataset can be used to efficiently train the GAN to dynamically generate the best MCS, repetitions, and PRB values. Moreover, the proposed MAB-based reinforcement learning mechanism helps overcome the lack of diversity in the generated samples in the GAN. The dataset generated by the MAB-based scheme contains values of the MCS, repetitions, and PRBs, which are dynamically selected considering different signal strengths. Thus, in the dataset, the diversity of the samples is maintained by the variation of the channel condition and dynamic adaptation of the aforementioned parameters. Therefore, at the time of the training of the GAN, the generator can generate samples by following the dynamics of the training dataset, and as a consequence, the lack of diversity in the generated samples is overcome.

\subsection{Contribution of this work}

%The proposed SmartCon is based on probabilistic deep learning governed by a point process mechanism that learns to schedule events with an adaptive selection of the MCS, repetitions and radio resources. 
%smart Industrial Internet of Things (IIoT)~\cite{da2014internet}
By exploiting online learning, the proposed model can provide an intelligent and unique NB-IoT framework for 5G and B5G networks. The main contributions of this work are summarized as follows:
\begin{enumerate}
 \item We design a GAN-based online learning model for auto link-configuration in NB-IoT. The model generates the real dynamics of the best possible MCS values, repetition numbers, and PRBs. Such dynamic adaptation targets to provide a low packet loss in the network.
 \item To generate the training dataset, we design a MAB-based reinforcement learning mechanism. It dynamically selects the aforesaid parameters by considering the present channel condition. As a result, a dataset is generated, that contains dynamic adaptation of MCS, repetitions, and PRB, that minimize the packet loss rate in the network.
The dataset is then used to train the GAN model. 
 \item %SmartCon is a medium access control (MAC) layer mechanism. 
For a thorough performance analysis, we implement a prototype of SmartCon in an NB-IoT compatible module of network simulator (NS) version NS-3 i.e., \texttt{ns-3-dev-NB-IOT}~\cite{ns3NBIoT}, by extending the LTE medium access control (MAC)~\cite{popli2019survey} module. The results show that SmartCon significantly improves the performance of NB-IoT systems compared to baselines.

%reduces the packet loss rate while improving other performance metrics, such as average throughput, delay, consumption of subframes, computational time, etc.
\end{enumerate}

\subsection{Organization of this paper}

The remainder of this paper is organized as follows. %Section~\ref{sec:overview} presents an overview of NB-IoT systems focusing on the architecture, and uplink and downlink framing structure. 
Section~\ref{sec:model} discusses the formulation of the TPP-based model to govern the propsoed GAN in SmartCon. The details of the proposed GAN model are described in Section~\ref{sec:gan}. The MAB-based mechanism used to generate the training dataset is discussed in Section~\ref{sec:MAB}. In Section~\ref{sec:impl}, the implementation details of SmartCon are presented along with details on the training mechanism. We analyze the performance of SmartCon in Section~\ref{sec:performance}, and Section~\ref{sec:concl} concludes this paper.

%\section{Background}

\section{TPP-based Model Formulation}
\label{sec:model}

%The proposed SmartCon is based on a deep probabilistic model influenced by point process.
In this section, we present the formulation of the TPP-based model that governs the propsoed GAN in SmartCon. 
%The proposed SmartCon is based on a deep probabilistic model influenced by point process. At the outset, SmartCon is a GAN~\cite{lei2019gcn} based framework, where the generative unit models – {\em the stochastic process of time-stamps of uplink and downlink scheduling of traffic, associated with allocated PRB, adaptive MCS and repetition number}. Whereas, the discriminative component regulates the processes performed by the generative model, such that the generator module can lead to the real dynamics which can be observed across several timestamps. 

\subsection{Time Series Modeling by Temporal Point Process}

A TPP is a stochastic process that contains isolated events at different time-stamps. Formally, a TPP is associated with a series of time-stamps $\mathcal{T}_t=\{t_l < t|l\in \mathbb{Z}^{+}\}$. Here, $\mathcal{T}_t$ denotes a set of occurrences of events which happened before time $t$. In the context of scheduling in NB-IoT, we define $\mathcal{T}_k(t)$ for eNB $k$ as the sequence of time instances of scheduling packet transmission (uplink or downlink) associated with the best possible MCS levels, repetition numbers, and the number of PRBs, based on the present channel condition, i.e., $\mathcal{T}_k(t)=\{t_l < t |\text{eNB}~k~\text{performs scheduling at time}~t_l\}$. Thus, $\mathcal{T}_k(t)$ is also called the history of scheduling conducted by eNB $k$ until time $t$. In addition, $\mathcal{T}_k(t)$ can also be expressed as a counting process defined by $N_k(t)\in \{0\}\cup \mathbb{Z}^{+}$, which keeps counting the number of scheduling operations in eNB $k$ during $[0,t)$. If $u(t-t_l)$ is a Heaviside step function, $N_k(t)$ can be represented as
%\begin{equation}\label{eq:count}
 $N_k(t)=\sum_{t_l\in \mathcal{T}_k(t)} u(t-t_l).$
%\end{equation}

Given the history $\mathcal{T}_k(t)$ of scheduling events until time $t$, we specify the dynamics of the counting process $N_k(t)$ using $\lambda_k(t)$ which captures the conditional probability of scheduling events associated with MCS levels, repetitions, and PRBs, in an infinitesimal time span $[t,t+dt)$. Let $dN_k(t)$ denote the number of such scheduling operations that are initiated by eNB $k$ in the time interval $[t,t+dt)$, and $dN_k(t)$ be equal to $1$. Thus, we have
%\begin{equation}\label{eq:lambda}
$\mathbb{P}(dN_k(t)=1|\mathcal{T}_k(t))=\lambda_k(t)dt.$
%\end{equation} 
We consider that scheduling occurrences are independent since the scheduling is influenced by the demand of packet transmission. Thus, we have 
%\begin{equation}\label{eq:count2}
 $\mathbb{P}(dN_k(t)=n|\mathcal{T}_k(t))=O(dt) \to 0 \quad \forall n \geqslant 2.$
%\end{equation}
Therefore, scheduling operations are asynchronous. So, $dN_k(t)$ can be $0$ or $1$, where $\lambda_k(t)$ needs to be considered when a scheduling occurs.
%by following probability $1$. 
Thus, we have
$$E[d\boldsymbol{N}(t)|\mathcal{T}_k(t)]=1.\lambda_k(t)dt + 0.(1-\lambda_k(t)dt)=\lambda_k(t)dt$$
\begin{equation}\label{countIntegration}
\text{i.e.} \quad E[d\boldsymbol{N}(t)|\mathcal{T}_k(t)]=\int_{0}^{T}\lambda_k(t)dt.
\end{equation}
Hence, $\lambda_k(t)$ also defines the average rate (intensity) of events which are occurring in an infinitesimal interval of time span $[t,t+dt)$. So, $\lambda_k(t)$ is also known as conditional intensity function, which may depend on $\mathcal{T}_k(t)$. It is noted that $\lambda_k(t)$ denotes the stochastic or random dynamics of $N_k(t)$. %Thus, the modeling of scheduling operations is mapped to the generation of $\lambda_k(t)$.

\subsection{Why Do We Need to Learn $\lambda_k(t)$ Instead of Applying a Parameterized Model?}

%In the majority of existing works, time-series modeling based on point process applies parameterized models from famous distributions like Hawkes process, Poisson process, etc. However, 
Parameterized distributions, such as Hawkes process, Poisson process, cannot capture the effects of various latent factors, such as the variation of signal strength and noise, on the real distribution of $\lambda_k(t)$. %; whereas, such latent factors are crucial for modeling wireless channels where packets are scheduled. 
For instance, the channel condition can affect the rate of packet transmission, while an inappropriate selection of MCS and repetition number can increase the packet loss rate and delay after scheduling. %Additionally, unavailability of packets in the buffer also makes a delay in scheduling. 
Therefore, by introducing such factors in the distribution of $\lambda_k(t)$, we learn the impact of the latent factors during scheduling. Next, we describe the proposed GAN. % this learning procedure. 

\subsection{The Reason of Applying GAN}

Considering the present channel condition, the packet loss rate and delay in the network depend on the MCS, repetitions, and PRBs selection. In our proposed model, the conditional intensity function $\lambda_k(t)$ represents the distributions of the stochastic time-stamps of traffic scheduling associated with the adaptive MCS, repetitions, and PRBs. In this context, we need to capture the effects of various latent factors (noise, interference, etc.) on the distribution of $\lambda_k(t)$ to learn the impact of the latent factors during scheduling. Since the GAN can generate the real dynamics by learning the patterns of data in the input dataset, the distribution of $\lambda_k(t)$ can be smartly modeled using the GAN. Consequently the stochastic time-stamps of traffic scheduling associated with the adaptive values of the aforementioned parameters can be intelligently generated.

\section{SmartCon: Modeling with GAN}
\label{sec:gan}

%A GAN~\cite{lei2019gcn} framework has a {\em generative module} and a {\em discriminative module}.
%The proposed SmartCon is based on a deep probabilistic model influenced by point process. 
%At the outset, SmartCon is a GAN~\cite{lei2019gcn} based framework, where the generative unit models {\em the stochastic time-stamps of uplink and downlink traffic scheduling associated with adaptive MCS values, repetitions, and number of PRBs}.
%Whereas, the discriminative component regulates the processes performed by the generative model, such that the generator module can lead to the real dynamics which can be observed across several time-stamps. 
%A GAN framework has a {\em generative module} and a {\em discriminative module}. %First, we describe the generator and the discriminator in our proposed GAN and then we discuss the overall MH-GAN model in our system.
%Particularly, 
The GAN module finds the distribution of $\lambda_k(t)$ by using the generative and discriminator modules, as shown in Fig.~\ref{fig:model1}. 
%The generative module produces the distribution of $\lambda_k(t)$ from the past observations of occurrences of events over time. In this case, a noise prior is included to generate $\lambda_k(t)$, and this noise helps capture the impact of silent factors, such as the variation of signal strength, poor channel condition, etc. The distribution of $\lambda_k(t)$ and the time-stamps are given as inputs to the discriminator module that helps the generative module produce the real dynamics of $\lambda_k(t)$ observed across several time-stamps by identifying fake distributions of $\lambda_k(t)$. %The real levels (ground truth levels) are compared with the fake levels on-the-go in order to fine tune the learning parameters associated with the discriminator. %At future time-stamps, 
%The real dynamics produced by the generator leads to the sequence of scheduling events associated with MCS levels, repetitions, and PRBs. 

\subsection{Generative Module}

\begin{figure}[!t]
\centering
\includegraphics[width=\linewidth]{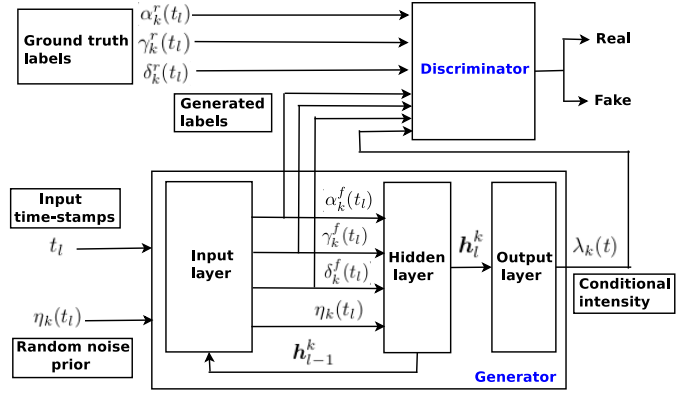}
%\vspace{-30mm}
\caption{Generator and discriminator modules in SmartCon}
\label{fig:model1}
%\vspace{-5mm}
\end{figure}

Let $\mathbb{K}$ be the set of all eNBs available in the network. For an eNB $k \in \mathbb{K}$, time-stamps of scheduling with MCS levels, repetition numbers, and PRBs are governed by the intensity function $\lambda_k(t)$, where this conditional intensity function generally depends on the past scheduling operations conducted by eNB $k$.  
%AP association for all APs $k\in \mathbb{K}$. 
We define $\lambda_k(t)$ as 
\begin{equation}\label{lambdaEqn}
\lambda_k(t)=\Upsilon(\mathcal{T}_k(t); \eta_k(t), \alpha_k(t), \gamma_k(t), \delta_k(t)).
\end{equation}
Here, $\Upsilon$ is an arbitrary nonlinear function which is modeled by a recurrent neural network (RNN), %It is a feedforward type artificial neural network containing some additional edges (called recurrent edges). 
%By using the hidden layers of the RNN, the last signals are connected to the present input. 
where the hidden layers help form recursive units which create an inbuilt memory. %; and consequently, RNNs are able to capture the influence of the past inputs on the present ones.
$\eta_k(t)$ is a seed variable or the noise prior, which is a usual input in deep generative models to capture the dynamics of the environment where the model is run. Specifically, in the proposed GAN, $\eta_k(t)$ introduces a variation of signal strength and noise. Along with $\eta_k(t)$, the proposed generative module is provided with three more sources of randomness -- $\alpha_k(t)$, $\gamma_k(t)$, and $\delta_k(t)$. These sources of randomness regulate the dynamics of transmission of the traffic components. All the random sources ($\eta_k(t)$, $\alpha_k(t)$, $\gamma_k(t)$, and $\delta_k(t)$) are instantiated only at the time-stamp where a packet is transmitted by the eNB. %, i.e., during timestamps $t_l\in \mathcal{T}(T)$.
These random sources are defined as follows:
\begin{enumerate}
 \item $\alpha_k(t)$: $\alpha_k\in\{0,1\}$ is a random variable that identifies the scheduling status of a packet at time $t$ in eNB $k$. When a packet is scheduled for an uplink or downlink transmission, the status is ON ($\alpha_k(t)=1$); otherwise, the status is OFF ($\alpha_k(t)=0$).

 \item $\gamma_k(t)$: When the scheduling status is ON for eNB $k$ (i.e., $\alpha_k(t)=1$), the number of PRBs used to transmit the packets scheduled at $t$ is determined by $\gamma_k(t)$. More specifically, $\gamma_k(t)\in [0,1]$ stores normalized values of the number of PRBs at time $t$ in eNB $k$.

%the probability of OWRP (i.e., $m^{'}_k(t)$) based on $\gamma_k(t)$. More specifically, $\gamma_k(t)\in [0,1]$ is a continuous random variable that indicates whether OWRP would occur or it would not occur with probabilities $\gamma_k(t)$ and $(1-\gamma_k(t))$, respectively.

 \item $\delta_k(t)$: This parameter is a pair of normalized values of MCS and repetition number, which are associated with the scheduling of a packet at time $t$ in eNB $k$. The value of $\delta_k(t)$ is defined when the scheduling status is ON i.e., $\alpha_k(t)=1$. 

%This parameter is a binary instantiation of $\gamma_k(t)$. So, $\delta_k(t)$ defines that OWRP would happen ($\delta_k(t)=1$) or it would not happen ($\delta_k(t)=0$). Thus, $\delta_k(t) \sim \text{Bernoulli} (\gamma_k(t))$. Table~\ref{table1} shows the relation between $\alpha_k(t)$ and $\delta_k(t)$.

%\begin{table}[h]
%\caption{Relation Between $\alpha_k(t)$ and $\delta_k(t)$}
%%\scriptsize
%\centering
%\begin{tabular}{|p{1cm}|p{1cm}|p{4cm}|}
%\hline
%\textbf{$\alpha_k(t)$} & \textbf{$\delta_k(t)$} & \textbf{Chance of OWRP} \\
%\hline
%\hline
%$0$ & $0$ & Channel bonding is disabled and OWRP would not occur  \\
%\hline 
%$1$ & $0$ & Channel bonding is enabled and OWRP would not occur \\
%\hline
%$1$ & $1$ & Channel bonding is enabled and OWRP would occur \\
%\hline
%\end{tabular} 
%\label{table1}
%\end{table}
\end{enumerate}
%Next, using the RNN, we design the distribution of the function $\lambda_k(t)$. 
In the RNN, recursive units help create an inbuilt memory, and thus the impacts of the past transmissions on the present transmission can be captured correctly. The proposed GAN uses one RNN ($\text{RNN}_k$) per eNB $k$. $\text{RNN}_k$ considers the previous time-stamps ($t_l\in \mathcal{T}_k(t)$) of scheduling packet transmissions associated with MCS levels, repetitions, and radio resources as inputs and generates the conditional intensity function $\lambda_k(t)$ for the scheduling events of the next packets. In this context, the hidden states of $RNN_k$ embed the history $\mathcal{T}_k(t)$ into the vectors $\boldsymbol{h}_l^k$ which are determined recursively by utilizing the previous information $\boldsymbol{h}_{l-1}^k$ and the signals acquired from the present input. For eNB $k$, such $\boldsymbol{h}_{\bullet}^k$ are fixed low dimensional representations of the history of scheduled packets associated with a MCS, repetition number, and number of PRBs.
Fig.~\ref{fig:model1} illustrates different parameters used in the generator, along with the discriminator module. 
%In the generator, the RNN has three layers which are {\em input layer}, {\em hidden layer}, and {\em output layer}. These three layers are modeled as follows.
In the generator, the RNN has three layers as follows.
%\noindent{\textbf{Input layer:}} 
\subsubsection{Input layer}

The activation of the input layer occurs when a packet is transmitted. Specifically, at the $l$-th transmission time ($t_l$), the input layer considers the previous states $\boldsymbol{h}_{l-1}^k$ as input and produces the random signals $\eta_k(t_l)$, $\alpha_k(t_l)$, $\gamma_k(t_l)$, and $\delta_k(t_l)$, which are fed into the next layer (the hidden layer). Particularly, at time-stamp $l\geqslant 1$, the input layer creates the aforesaid random signals as follows.
\begin{itemize}
 \item \textbf{Definition of $\eta_k(t_l)$:} At time-stamp $t_l$, a Poisson distribution is used to generate the noise prior, i.e., $\eta_k(t_l) \sim \text{Poisson} (\mu)$, where $\mu\geqslant 0$ is average number of occurences of events per interval. % and $\mu\geqslant 0$.

 \item \textbf{Definition of $\alpha_k(t_l)$:} The random variable, $\alpha_k(t_l)$, which decides whether packet scheduling is ON/OFF is sampled from a Bernoulli distribution. The mean of this distribution is represented by a logistic function of the preceding hidden state $\boldsymbol{h}_{l-1}^k$. That is, 
   \begin{equation}
     \alpha_k(t_l)=\text{Bernoulli} (\xi_k (t_l)),
   \end{equation}
  where $\xi_p (t_l)=\sigma(\boldsymbol{w}_{\alpha}^T \boldsymbol{h}_{l-1}^k)$. Here, $\boldsymbol{h}_{l-1}^k$ is the output of the hidden layer, which represents the state of the RNN at time-stamp $t_{l-1}$. When $\boldsymbol{h}_{0}^k=0$, $\alpha_k(t_l)=\text{Bernoulli} (1/2)$.

  \item \textbf{Definition of $\gamma_k(t_l)$:} The density function of $\gamma_k(t_l)$ depends on the noise $\eta_k(t_l)$, and thus $\gamma_k(t_l)$ is defined using a standard normal distribution, as follows.
   \begin{equation}\label{eq:gamma}
     \gamma_k(t_l)=\frac{1}{\sqrt{2\pi}} exp\Big(-\frac{(\eta_k(t_l))^2}{2}\Big)
   \end{equation}

  \item \textbf{Definition of $\delta_k(t_l)$:} In eNB $k$, at time-stamp $t_l$, let $m_k(t_l)$ and $r_k(t_l)$ be random variables that represent the MCS and repetition number, respectively. Since the selection of MCS values and repetition numbers is influenced by $\eta_k(t_l)$, $m_k(t_l)$ and $r_k(t_l)$ are defined as  
   \begin{equation}\label{eq:mcs}
     \mathbb{P}(m_k(t_l)|\alpha_k(t_l)=1)= \frac{\beta exp(-\beta \eta_k(t_l))}{1-exp(-\beta)},
   \end{equation}
   \begin{equation}\label{eq:rn}
     \mathbb{P}(r_k(t_l)|\alpha_k(t_l)=1)= \beta exp(-\beta \eta_k(t_l)).
   \end{equation}
Eqns.~(\ref{eq:mcs}) and~(\ref{eq:rn}) indicate that $m_k(t_l)$ and $r_k(t_l)$ follow exponential distributions and take values between $[0,1]$. Eqns.~(\ref{eq:mcs}) and~(\ref{eq:rn}) allow us to generate random variables that follow the exponential distribution and depend on another random variable. Since $\eta_k(t)$ introduces a variation in signal strength and noise, the MCS $m_k(t_l)$ and repetition number $r_k(t_l)$ depend on $\eta_k(t)$, and therefore $m_k(t_l)$ and $r_k(t_l)$ are calculated using Eqns.~(\ref{eq:mcs}) and~(\ref{eq:rn}). In this context, we consider exponential distributions for $m_k(t_l)$ and $r_k(t_l)$ because their impacts on the network performance are significantly influenced by the variation of signal strength and noise. In Eqns.~(\ref{eq:mcs}) and~(\ref{eq:rn}), a difference is added to the denominator to impose a variation between the values of $m_k(t_l)$ and $r_k(t_l)$. The MCS and repetition number are selected when a packet is scheduled, and therefore values of the MCS and repetition number are defined when $\alpha_k(t_l)=1$. The functional forms of Eqns.~(\ref{eq:mcs}) and~(\ref{eq:rn}) are borrowed from~\cite{rolfe2016discrete}.
In Eqns.~(\ref{eq:mcs}) and~(\ref{eq:rn}), $\beta>0$. Since $\delta_k(t_l)$ is a pair of values, we define $\delta_k(t_l)$ as 
   %\begin{equation}\label{eq:delta}
    $ \delta_k(t_l) = \{m_k(t_l),r_k(t_l)\}.$
   %\end{equation}
When $\alpha_k(t_l)=0$, scheduling is not performed. Thus, $\delta_k(t_l)=0$ is deterministic when $\alpha_k(t_l)=0$. Therefore, we have
   \begin{equation}\label{eq:delta2}
     \mathbb{P}(\delta_k(t_l)|\alpha_k(t_l)=0)= \text{Dirac delta} (\delta_k(t_l)).
   \end{equation}
The functional form of Eqn.~(\ref{eq:delta2}) is borrowed from~\cite{rolfe2016discrete}.

% \item The density function of $\gamma_k(t_l)$ is defined as follows.
%   \begin{equation}\label{eq1}
%     \mathbb{P}(\gamma_k(t_l)|\alpha_k(t_l)=1)= \frac{\beta exp(-\beta \gamma_k(t_l)}{1-exp(-\beta)}
%   \end{equation}
%   \begin{equation}\label{eq2}
%     \mathbb{P}(\gamma_k(t_l)|\alpha_k(t_l)=0)= \text{Dirac delta} (\gamma_k(t_l))
%   \end{equation}
% Eqn.~(\ref{eq1}) indicates that $\gamma_k(t_l)$ follows a distribution that is exponential and belongs to the range of $[0,1]$. Whereas, Eqn.~(\ref{eq2}) shows that $\gamma_k(t_l)$ is deterministic with the value of $0$ when $\alpha_k(t)=0$. The functional forms of Eqn.~(\ref{eq1}) and Eqn.~(\ref{eq2}) are borrowed from~\cite{rolfe2016discrete}.
  
  %\item Finally, $\delta_k(t_l) \sim \text{Bernoulli}(\gamma_k(t_l))$. Thus, $\delta_k(t_l)=1$ indicates that OWRP would happen and $\delta_k(t_l)=0$ signifies that OWRP would not occur. 

\end{itemize} 

%\noindent{\textbf{Hidden layer:}} 
\subsubsection{Hidden layer}

The input time-stamps $t_l$ and the random signals produced in the previous layer are used to create the next state $\boldsymbol{h}_{l}^k$ based on the present hidden state $\boldsymbol{h}_{l-1}^k$. The definition of $\boldsymbol{h}_{l}^k$ is 
\begin{multline}\label{eq:hiddenstate}
 \boldsymbol{h}_{l}^k = \Omega_g\Big(\boldsymbol{W}_1 \boldsymbol{h}_{l-1}^k+ \boldsymbol{W}_2 \alpha_k(t_l)\big(\gamma_k(t_l)+\delta_k(t_l)\big)+\\
 \boldsymbol{W}_3\big(1-\alpha_k(t_l)\big)t_l\eta_k(t_l)+\boldsymbol{b_h} \Big). 
\end{multline}
Here, $\Omega_g$ is an activation function, and $\boldsymbol{W}_1$, $\boldsymbol{W}_2$, $\boldsymbol{W}_3$ and $\boldsymbol{b_h}$ are trainable parameters. $\Omega_g$ uses the {\em Rectified Linear Unit (ReLU)} activation function, which requires less computations than other activation functions. 
To overcome the vanishing gradient problem, we use the {\em Rectified Linear Unit (ReLU)} as the activation function in the hidden layers~\cite{shin2016predicting}. The ReLU does not cause a small derivative. When the value of the input variable is greater than $0$, the gradient of the ReLU is $1$, and zero otherwise. Therefore, multiplying a set of ReLU derivatives in the backpropagation equations results in $0$ or $1$, and consequently, there is no `vanishing' of the gradient.

Note that the proposed model is stateful, which is a key distinguishing characteristic. Normalized values of the number of PRBs, MCS and repetition number need to be considered when a packet is scheduled for uplink or downlink transmission, i.e., the value of $\alpha_k(t_l)$ is $1$. Thus, in Eqn.~(\ref{eq:hiddenstate}), $\gamma_k(t_l)$ and $\delta_k(t_l)$ are multiplied by $\alpha_k(t_l)$, along with the trainable parameter $\boldsymbol{W}_2$. When packet scheduling is not performed, the MCS and repetition number are not required, and therefore only the noise value is considered with the time instant. This scenario is represented by the term $\boldsymbol{W}_3\big(1-\alpha_k(t_l)\big)t_l\eta_k(t_l)$, with the trainable parameter $\boldsymbol{W}_3$.

%\noindent{\textbf{Output layer:}} 
\subsubsection{Output layer}

Based on the hidden states, the output layer generates the conditional intensity $\lambda_k(t)$ as
\begin{equation}\label{eq:intensity}
 \lambda_k(t) = exp(\boldsymbol{W_g}^T \boldsymbol{h}^k_l + \boldsymbol{c_g}(t-t_l) + \boldsymbol{b_g}).
\end{equation}
Here, $t_l<t$ and $\lambda_k(t)$ samples the next time-stamp by applying Ogata's thinning algorithm~\cite{ogata1981lewis}. Let $\theta_G=\{\boldsymbol{W}_1, \boldsymbol{W}_2, \boldsymbol{W}_3,\boldsymbol{b_h},\boldsymbol{W_g},\boldsymbol{c_g},\boldsymbol{b_g} \}$ be trainable parameters used in the generative model. Under the generative framework, the log-likelihood of $\lambda_k(t)$ can be defined as 
\begin{equation}
 \log \mathcal{L}(\lambda_k|\theta_{G}) = \sum_{j=1}^{|\mathcal{H}(T)|} \log \lambda_k(t_l) - \int_{0}^{T}\lambda_k(t)dt.
\end{equation}
 
\subsection{Discriminative Module}

%Let the tuple $(t_l,\alpha_k^{*}(t_l),\gamma_k^{*}(t_l),\delta_k^{*}(t_l))$ denote that $\alpha_k(t_l)$, $\gamma_k(t_l)$ and $\delta_k(t_l)$ related information is transmitted at time $t_l$, and it has values $\alpha_k^{*}(t_l)$, $\gamma_k^{*}(t_l)$ and $\delta_k^{*}(t_l)$. 
%The $\alpha_k(t_l)$, $\gamma_k(t_l)$, and $\delta_k(t_l)$ related information which are fed into the discriminator may be fake or real values. 
In general, let $\alpha_k^{*}(t_l)$, $\gamma_k^{*}(t_l)$, and $\delta_k^{*}(t_l)$ be the values fed into the discriminator, which may be fake or real.
The discriminative unit takes a series $F$ of fake data generated by the generative module and a series $R$ of real (observed) values for $F$. Specifically, we represent $F$ and $R$ as $F=(\alpha_k^{f}(t_l),\gamma_k^{f}(t_l),\delta_k^{f}(t_l))$ and $R=(\alpha_k^{r}(t_l),\gamma_k^{r}(t_l),\delta_k^{r}(t_l))$. %Note that $\delta_k^{r}(t_l)$ is binary instantiation of $m^{'}_k(t_l)$ (discussed in Section~\ref{sec:OWRPrep}). 
We design the discriminator using an RNN whose hidden layer for eNB $k$ at time $t_l$ is defined in what follows.
\begin{multline}\label{eq:dis_hidden}
 \Phi_{l}^{k}\big(\alpha_k^{*}(t_l),\gamma_k^{*}(t_l),\delta_k^{*}(t_l) \big)= \Omega_d \Big(\boldsymbol{W}_4 \Phi_{l-1}^{k} + \boldsymbol{W}_5\big(\alpha_k^{*}(t_l)+ \\ 
  \alpha_k^{*}(t_l)\gamma_k^{*}(t_l)\delta_k^{*}(t_l)\big) + \boldsymbol{b_d} \Big)
\end{multline}
At each time $t_l$, the hidden layer of the discriminative model outputs $\Phi_{l}^{k}$, which defines the probability of correctness of $\alpha_k^{*}(t_l)$, $\gamma_k^{*}(t_l)$, and $\delta_k^{*}(t_l)$, i.e., if they belong to $R$. 
In Eqn.~(\ref{eq:dis_hidden}), $\Omega_d$ is the sigmoid activation function. From~(\ref{eq:dis_hidden}), it is noted that $\gamma_k^{*}(t_l)$ and $\delta_k^{*}(t_l)$ have no effect when $\alpha_k^{*}(t_l)$ is zero. This protects against noise in the input data, where $\gamma_k^{*}(t_l)$ and $\delta_k^{*}(t_l)$ are non-zero while $\alpha_k^{*}(t_l)$ is zero. Assume that $\theta_{D}=\{\boldsymbol{W}_4,\boldsymbol{W}_5,\boldsymbol{b_d}\}$ are the trainable parameters for the discriminator. In case of real sequence $R$, the log-likelihood of the discriminator (expected value of $\log D_{\theta_{D}}$) is defined as
\begin{equation}\label{eq:real}
 E_{R,\theta_{D}}[\log D_{\theta_{D}}]  = \sum_{j=1}^{|R|}\log \Phi_{l}^{k}\big(\alpha_k^{r}(t_l),\gamma_k^{r}(t_l),\delta_k^{r}(t_l) \big).
\end{equation}
For a fake sequence ($F$), the log-likelihood of the discriminator is
\begin{multline}\label{eq:fake}
 E_{R,\theta_{G},\theta_{D}}[\log (1-D_{\theta_{D}})]  = \sum_{j=1}^{|F|}\log \Big(1-\Phi_{l}^{k}\big(\alpha_k^{f}(t_l),\\
  \gamma_k^{f}(t_l),\delta_k^{f}(t_l) \big)\Big).
\end{multline}

Now, in SmartCon, the loss function of the proposed GAN model is defined as 
\begin{multline}\label{eq:objective}
 \min\limits_{\theta_{G}}\max\limits_{\theta_{D}}~ -\log \mathcal{L}(\lambda_k|\theta_{G})+E_{R,\theta_{D}}[\log D_{\theta_{D}}]+ \\
   E_{R,\theta_{G},\theta_{D}}[\log (1-D_{\theta_{D}})].
\end{multline}
Therefore, SmartCon maximizes the log-likelihood of the conditional intensity $\lambda_k$ and optimizes the adversarial objective for generating the labels ($\alpha_k^{*}(t_l)$, $\gamma_k^{*}(t_l)$, and $\delta_k^{*}(t_l)$). %Fig.~\ref{fig:model1} illustrates the generator and the discriminator modules. 
At a time, only one data sample is processed in the stochastic gradient descent (SGD), and thus the SGD is computationally fast. In addition, since the SGD causes more frequent updates to the parameters, it has faster convergence for larger datasets~\cite{bottou201113}. Therefore, the SGD is used to solve the optimization problem in Eqn.~(\ref{eq:objective}).

\subsection{Learning with GAN}

Once the GAN model can generate the real labels' dynamics ($\alpha_k^{*}(t_l)$, $\gamma_k^{*}(t_l)$, and $\delta_k^{*}(t_l)$), SmartCon performs the predictions described in what follows.
\begin{itemize} 
 \item Based on $\alpha_k^{*}(t_l)$, the eNB predicts the probability of packet scheduling at time $t_l$.
 \item Based on $\gamma_k^{*}(t_l)$, the required number of PRBs for the scheduling is chosen.  
 \item Based on $\delta_k^{*}(t_l)$, eNB $k$ selects the best possible MCS level and repetition number for the scheduling at $t_l$.
\end{itemize}

In particular, the GAN does not belong to the traditional reinforcement learning model. However, considering a reinforcement learning approach, the proposed GAN has three {\em states} -- (i) the generation of $\alpha_k^{f}(t_l)$, $\gamma_k^{f}(t_l)$, and $\delta_k^{f}(t_l))$, by the generator, (ii) the generation of $\lambda_k(t)$ by the generator, and (iii) the differentiation between the fake and real values of $\alpha_k(t_l)$, $\gamma_k(t_l)$, and $\delta_k(t_l))$, by the discriminator. The optimization function in Eqn. (14) can be considered as the {\em reward}. The action space can be defined as a set of actions that transfer the data produced in a state to another state of the GAN. 

%% Training of the Model
\section{Training Dataset Generation}
\label{sec:MAB}

%In this section, we discuss the mechanism used to generate the training dataset in order to dynamically select the best MCS values, repetitions, and radio resources, for transmissions during uplink or downlink scheduling, such that the PLR is reduced. This adaptive selection is based on the MAB online learning approach given the present SINR value of the channel. The scheduled packets are transmitted using the selected MCS level, repetition, and number of PRBs. Then, the PLR obtained after the transmission is stored for future reference. %We apply MAB-based online learning approach to dynamically select MCS and repetition number. 
%The selected MCS, repetition, and number of radio resources serve as the training dataset for the proposed GAN model in SmartCon.
%In this section, we first discuss the MAB-based learning mechanism and then present the application of MAB for the dynamic selection of MCS, repetitions, and radio resources, in order to prepare the training dataset. 

In this section, we present our MAB-based~\cite{Robbins:1952} dynamic selection of MCS, repetitions, and radio resources, in order to prepare the training dataset. Specifically, the $\epsilon$-greedy algorithm which is a variant of MAB mechanism is used in our proposed mechanism.

%\subsection{Multi-Armed Bandit}

%%In MAB models~\cite{Robbins:1952}, the objective is to balance the trade-off between {\em exploration} and {\em exploitation} by using a learning agent. 
%In MAB models~\cite{Robbins:1952}, a learning agent opts for a single option (known as {\em arm}) from a set of available options which have unknown characteristics at the initial stage. Based on its choice, a certain {\em reward} is received by the agent. %Therefore, the agent chooses an arm and gets a reward for it. 
%The agent always tries to maximize the cumulative reward. %% obtained over a given period of time. %Originally, the bandit problem was presented by Robbins (1952)~\cite{Robbins:1952}. 
  
\subsection{$\epsilon$-greedy Algorithm}

%The $\epsilon$-greedy policy~\cite{Epsilon-greedy} is a popular variant of the MAB model, and 
We use $\epsilon$-greedy policy~\cite{Epsilon-greedy} as a MAB mechanism to dynamically select the MCS and repetition number given the present signal-to-interference-plus-noise ratio (SINR) value of the channel. The $\epsilon$-greedy mechanism uses a parameter $\epsilon$ as exploration probability. At time instant $t$, $\epsilon_t$ is defined as 
%\begin{equation} \label{epsilon}
$\epsilon_t=min(1, cK/d^2t).$
%\end{equation}
Here, $K$ is the total number of arms used in the bandit problem. The parameter $c\geq 0$ is a small integer. The parameter $d$ specifies the difference between the expected rewards of the best and second best arms. Here, the best arm denotes the arm that has provided the maximum average reward so far. %, and the second best arm gives the next highest expected reward. 
The $\epsilon$-greedy policy is described by two phases listed below. 

\begin{itemize}
  \item \textbf{Exploration:} In the exploration phase, we randomly choose an arm from the available set of arms. The probability of exploration is defined by $\epsilon$.
  \item \textbf{Exploitation:} In the exploitation phase, we choose the arm associated with the maximum average reward so far. In this case, $(1 - \epsilon)$ defines the probability of exploitation.
\end{itemize} 
%Therefore, the {\em strategy} of the $\epsilon$-greedy policy can be represented as strategy = $\epsilon \times$ explore + $\big (1 - \epsilon \big ) \times$ exploit. 
In~\cite{Auer:2002}, it is described that after $n$ number of plays, the probability of choosing a suboptimal arm is upper bounded by ${O}(c/d^2 n)$, where $n \ge cK/d^2$. 

%Next, we detail the MAB-based learning approach. %to adaptively select the best possible MCS and repetition number, based on the present SINR value of the channel. 

\subsection{Exploiting MAB for Dynamic Selection of MCS and Repetition Number}

Let $\mathcal{M}$ be the set of available MCS values, $\mathcal{R}$ the set of available repetition numbers, and $\mathcal{P}$ the set of number of PRBs available in NB-IoT systems. Let $\mathcal{M}=\{M_1,M_2,M_3,...,M_p\}$, $\mathcal{R}=\{R_1,R_2,R_3,...,R_q\}$, and $\mathcal{P}=\{P_1,P_2,P_3,...,P_u\}$, where $p>0$, $q>0$, and $u>0$ are the counts of the available MCS values, repetitions, and PRBs, respectively. In our MAB model, the selection of MCS levels, repetitions, and number of PRBs is the {\em arm} and we refer to it as {\em MCS-Repetition-PRB (M-R-P) configuartion}. Let $\mathcal{A}$ be the arm, and therefore the arm with $a^{th}$ MCS value, $b^{th}$ repetition number, and $c^{th}$ number of PRBs can be represented as $\mathcal{A}_{abc}=\{M_a,R_b,P_c\}$, where $1\le a\le p$, $1\le b\le q$ and $1\le c\le u$. Thus, $K$ specifies the total count of $\mathcal{A}_{abc}$. In the dynamic selection of MCS values, repetitions, and PRBs, the objective is to minimize the packet loss rate after scheduling. Thus, in our MAB model, the reward is the inverse of the PLR and let $\mathcal{D}$ denote PLR. %Therefore, $\text{reward}=1/\mathcal{D}$. %Therefore, we have $\mathcal{R}=1/\mathcal{D}$. 
%Hence, a lower value of the PLR leads to a higher reward. 

\subsubsection{Statistic table}

We use a statistic table, denoted by $\mathbb{S}=\{\mathcal{S},\mathcal{A},\mathcal{D}\}$, in order to store information regarding the selected M-R-P configuration for the present SINR of the channel. $\mathbb{S}$ also stores the PLR observed against the values of aforesaid selected parameters. %Additionally, this table also stores the number of PRB used to transmit all the scheduled packets. 
%In general, $\mathbb{S}$ can be represented as $\mathbb{S}=\{\mathcal{S},\mathcal{A},\mathcal{D}\}$. 
$\mathcal{S}$ denotes the SINR of the channel.
%and $\mathcal{N}_{PRB}$ denotes the number of PRB, which is used under $\mathcal{A}$ and $\mathcal{S}$.

\subsubsection{Execution of the MAB Approach}

%The mechanism for the dynamic selection of the MCS, repetitions, and PRBs, which is executed by a learning agent $\mathcal{L}$, is given in Algorithm~\ref{algo1}. %Details of the mechanism are discussed as follows.
In Algorithm~\ref{algo1}, there are two stages -- {\em (1) initial stage}, and {\em (2) experience stage}. %In the initial stage, the MAB model explores some of the M-R-P configurations from the set of available configurations. %In the experience stage, the algorithm is executed to select the best possible M-R-P configuration. 
Descriptions of these two stages are given in what follows.

\noindent {\textbf{(1) Initial stage}}: The learning agent $\mathcal{L}$ calculates the SINR of the channel and selects the M-R-P configuration randomly from the set of available configurations. After a time period of $t_{d}$, $\mathcal{L}$ calculates the PLR and computes the reward accordingly. %This enables $\mathcal{L}$ to gather information about the wireless environment by choosing arbitrary MCS, repetitions, and number of PRBs. $\mathbb{S}$ is updated at the end of each $t_{d}$. 
Therefore, the initial stage helps the agent populate $\mathbb{S}$ to start the experience stage.

\noindent{\textbf{(2) Experience stage:}} %In this phase, both exploitation and exploration are performed. 
The description of the exploitation is as follows. 

\noindent{\textbf{Exploitation:}} At time $t$, let the SINR be $S_t$ and the exploitation be executed with probability $(1-\epsilon_t)$. %In case of exploitation, we consider two scenarios -- (i) \textit{Case-1:} the present SINR value is found in $\mathbb{S}$ and (ii) \textit{Case-2:} the current SINR value is not present in $\mathbb{S}$. %At time $t$, these two cases are executed with probability $(1-\epsilon_t)$. 
%The first allows the exploitation of past knowledge to select the best M-R-P configuration for the present SINR. Whereas, the second case uses the best past experience without considering the present SINR since the current SINR is not found in $\mathbb{S}$. This approach fulfills the objective of the exploitation scheme, i.e., the application of the best configuration. 
%At time $t$, let the SINR be $S_t$. %The following selections are performed with probability $(1-\epsilon_t)$.
%Details of these two cases are provided in what follows.
We consider two scenarios as follows.
%\begin{itemize}

\begin{enumerate}
   \item \textbf{Case-1} (\textit{Consideration of a subset of $\mathbb{S}$:}) This case allows the exploitation of past knowledge to select the best M-R-P configuration for the present SINR. In this context, a small value $\Delta > 0$ is chosen to define the range of the SINR in $\mathbb{S}$, where the present SINR is found. %Therefore, a subset of the  $\mathbb{S}$ is considered and let $\mathbb{S}^S$ denote this subset. 
Hence, specifically, {\em Case-1} can be defined as follows.
If $S_{t} \in [(S_{t} - \Delta),(S_{t} + \Delta)]$ in $\mathbb{S}^S \subset \mathbb{S}$, the M-R-P configuration $\mathcal{A}_t$ is chosen from $\mathbb{S}^S$ such that $\mathcal{A}_t$ provides the lowest PLR in the set $\mathbb{S}^S$.

   \item \textbf{Case-2} (\textit{Consideration of the entire $\mathbb{S}$:}) 
The second case uses the best past experience without considering the present SINR since it is not found in the range of the SINR defined by $\Delta$ in $\mathbb{S}$. %This approach also fulfills the objective of the exploitation scheme, where the best M-R-P configuration is selected from the whole $\mathbb{S}$.
Particularly, {\em Case-2} is defined as follows.
If $S_{t} \notin [(S_{t} - \Delta),(S_{t} + \Delta)]$ in $\mathbb{S}$, the M-R-P configuration $\mathcal{A}_t$ is chosen from the entire $\mathbb{S}$ such that $\mathcal{A}_t$ provides the lowest PLR in $\mathbb{S}$.
\end{enumerate}

\noindent{\textbf{Exploration:}} %The exploration is performed with probability $\epsilon_t$. 
An M-R-P configuration is selected randomly with probability $\epsilon_t$ from the M-R-P configuration. %Exploration needs to happen to explore performances of M-R-P configurations that have not been visited yet. Otherwise, many configurations that could provide lower PLRs than the explored configurations, will remain unexplored.

%In the exploitation and exploration, after the transmission of all the scheduled packets, $\mathbb{S}$ is updated with $\mathcal{S}$, $\mathcal{A}$, and $\mathcal{D}$.
%The statistic table is updated after the exploitation and exploration.

\begin{algorithm}[!t]
\caption{MAB-based Selection of MCS Levels, Repetitions and PRBs}
\label{algo1}
\begin{algorithmic}[1]
\scriptsize
\State \textbf{Start}
\State \textbf{Initial stage:} Calculate the present SINR $S_t$ of the channel and select the M-R-P configuration randomly from the set of available configurations. After a time period $t_{d}$, compute the packet loss rate.
\State \textbf{Experience stage:} At time $t$, calculate the present $S_i$.
\State Calculate $\epsilon_t$ by using $\epsilon_t=min(1, cK/d^2t)$.
\State Let $\zeta \leftarrow $ Random(0,1).
\If{$\zeta \leq \epsilon_t$}
 \If{$S_{t} \in [(S_{t} - \Delta),(S_{t} + \Delta)]$ in $\mathbb{S}^S \subset \mathbb{S}$}
   \State Choose M-R-P configuration $\mathcal{A}_t$ from $\mathbb{S}^S$ such that $\mathcal{A}_t$ provides the lowest PLR (i.e. $\mathcal{D}$) in $\mathbb{S}^S$.
 \Else
   \State Select M-R-P configuration $\mathcal{A}_t$ from $\mathbb{S}$ such that $\mathcal{A}_t$ provides the lowest PLR in $\mathbb{S}$.
 \EndIf
 \Else
   \State Choose an M-R-P configuration at random from the available set of M-R-P configurations.
\EndIf
\State $\mathcal{D}$ is calculated.
\State $\mathbb{S}$ is updated with $\mathcal{S}$, $\mathcal{A}$, and $\mathcal{D}$.
%\State Update $\mathbb{S}$.
\State \textbf{End}
\end{algorithmic}
\end{algorithm}

\section{Implementation and Training Details}
\label{sec:impl}

%The LTE module inNS-3 includes aspects such as radio resource management (RRC), physical layer error model[15], QoS-aware packet scheduling, inter-cell interference coordination, and dynamic spectrumaccess. Based on the LTE module in NS-3, the authors of [12] implemented the basic featuresfor eMTC and NB-IoT modules. Based on the NB-IoT module described in [12], we implementthe uplink coverage enhancement features.
We implement SmartCon in \texttt{ns-3-dev-NB-IOT}~\cite{ns3NBIoT} with one eNB, where the number of UEs is varied from $10$ to $100$.
%In this section, we present the implementation of SmartCon in \texttt{ns-3-dev-NB-IOT}~\cite{ns3NBIoT}. 
The NB-IoT module belongs to LTE Cat NB1, where the downlink and uplink peak data rates are $26$ kbps and $66$ kbps, respectively. Both uplink and downlink transmissions are considered. The NB-IoT module in NS-3 includes numerous features, such as radio resource control (RRC), radio link control (RLC), packet scheduling, physical layer error model, inter-cell interference coordination, dynamic spectrum access, etc.~\cite{rico2016overview,malik2018radio}. We have used these aspects in the implementation of our proposed mechanism. We vary the levels of the interference in order to analyze the performance of SmartCon in different channel conditions. We consider both UDP and TCP packets with a ratio of $80\%$ and $20\%$, respectively. We use {\em proportional fair scheduling} to schedule the packets. The UEs are placed following a Poisson distribution centered at the eNB's position. To set the MCS, TBS, PRB, and code rate for a channel condition, we have applied the standard tables defined by the 3GPP standard~\cite{3GPPIoT}. Unless stated otherwise, we set the number of UEs to $100$. The SINR is chosen randomly between $5$dB--$25$dB. Details on the simulation setup are given in Table~\ref{table:sim}.

\begin{table}[!t]
\scriptsize
%\tiny
\caption{Simulation Parameters}
\centering
\begin{tabular}{|p{3.3cm}|p{4.2cm}|}
\hline
\textbf{Parameter} & \textbf{Value}\\
\hline 
%Adaptive Modulation and Coding (AMC) model & Vienna \\
%\hline
%\hline
Frequency Band & DL: $925$ MHz, UL: $880$ MHz\\
\hline
Default Transmission Mode & $0$ (Single-input-single-output (SISO))\\
\hline
Path loss model & FriisSpectrumPropagationLossModel \\
\hline
Fading model & TraceFadingLossModel \\
\hline
Propagation model & Okumura-Hata (Open area), Hybrid building(Urban)\\
\hline
NoiseFigure of UE & $5$ dB\\
\hline
NoiseFigure of eNB & $9$ dB \\
\hline
Downlink peak data rate & $26$ kbps\\
\hline
Uplink peak data rate & $66$ kbps\\
\hline
%Maximum physical data rate & $250$kbps \\
%\hline
Propagation delay model & Constant speed propagation delay model\\
\hline
Bit error rate (BER) & $0.03$ \\
\hline
UE scheduler type & PfFfMacScheduler \\
\hline
Packet Size & $100$ bytes \\
\hline
Mobility model & Random direction 2d mobility model (``Bounds: Rectangle (-100, 100, -100, 100)'', ``Speed: ConstantRandomVariable [Constant=3.0]'', ``Pause: ConstantRandomVariable [Constant=0.4]'')\\
\hline 
System bandwidth & $180$ kHz\\
\hline
TxPower of UE & $23$ dBm \\
\hline
TxPower of eNB & $46$ dBm \\
\hline
Cell radius & $1.5$ km\\
\hline
Transmission mode & Multi-Tone \\
\hline
Receiver Chains & $1$ SISO\\
\hline
Number of Antennas & $1$\\
\hline
Duplex Mode & Half duplex \\
\hline
\end{tabular} 
\label{table:sim}
%\vspace{-5mm}
\end{table}

\subsection{Baseline Mechanisms}

%In order to analyze the performance of SmartCon, 
We have considered NANIS~\cite{yu2020npdcch} and NBLA~\cite{yu2017uplink} as baselines along with the standard scheduling approach in NB-IoT. NANIS addresses the adaptation problem of the time interval between two consecutive NPDCCHs. %In addition, NANIS minimizes the radio resources used to receive data in downlink transmissions. %Primarily, NANIS attempts to apply as many NPDSCH subframes as possible. 
NBLA is a threshold-based approach, where an uplink link adaptation is performed with the determination of the MCS value and repetition number. %based on the current channel condition. 
%In addition, the transmission gap and power control related uplink scheduling technologies are analyzed in NBLA. %However, this work is a threshold-based approach considering only the downlink scheduling in NB-IoT. 
The standard approach is basically a First-In First-Out (FIFO) mechanism with a static MCS value and no repetition number. Here, we set the MCS to $6$.
%In this case, we consider an average value of $6$ as the static MCS since it is sustainable in both low and high signal strengths. 
We also compare the performance of SmartCon with GAN-powered deep distributional Q network (GAN-DDQN) to add a comparison with a mechanism that combines the GAN and reinforcement learning. However, the GAN-DDQN is a dynamic allocation mechanism of network slicing resources in 5G communications.  

\subsection{Implementation of SmartCon}

\begin{figure}[!t]
 \centering
 \includegraphics[width=\linewidth]{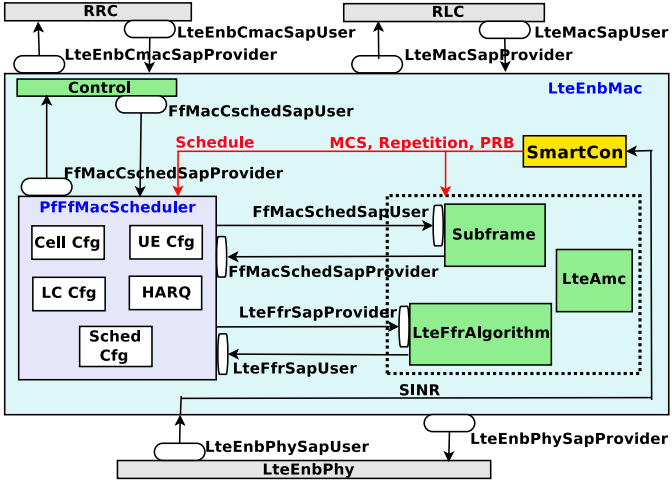}
 \caption{SmartCon implementation modules in \texttt{ns-3-dev-NB-IOT}}
 \label{fig:NB-PTS-impl}
 %\vspace{-5mm}
\end{figure}

We have implemented SmartCon by extending the LTE MAC~\cite{popli2019survey} module of \texttt{ns-3-dev-NB-IOT}, as shown in Fig.~\ref{fig:NB-PTS-impl}. The \texttt{OnOffApplication} is used to generate the traffic. The MAC layer functionalities of the eNB are implemented by the class \texttt{LteEnbMac} and \texttt{PfFfMacScheduler} implements the {\em proportional fair scheduler} to perform scheduling of UEs. In \texttt{LteEnbMac}, there are five interfaces for handling subframe, control information, packet scheduling (uplink and downlink), MCS assignment, and PRB allocation, implemented by a subframe block, control block, scheduler block, \texttt{LteAmc}, and \texttt{LteFfrAlgorithm}, respectively. Specifically, \texttt{LteFfrAlgorithm} is the base class that allocates PRBs and subframes for the transmission of data using a frequency reuse algorithm. %\texttt{LteAmc} implements an Adaptive Modulation and Coding Scheme (AMC) for the assignment of a modulation and coding rate.
%\texttt{LteFfrSapProvider} and \texttt{LteFfrSapUser} are the Service Access Points (SAPs) that correspond to the \texttt{LteFfrSap} interface. %The MAC scheduler communicates with the control block through \texttt{FfMacCschedSapProvider} and \texttt{FfMacCschedSapUser}. 
%To perform data transmission, \texttt{FfMacSchedSapUser} and \texttt{FfMacSchedSapProvider} help establish a connection between the subframe block and MAC scheduler. 
%In the MAC scheduler, the interfaces \texttt{LteMacSap}, \texttt{LteEnbCmacSap}, and \texttt{LteEnbPhySap} establish the communications with the RLC, RRC, and physical (PHY) layer, respectively. %The MAC scheduler holds several blocks. The \textit{HARQ} block is used for Hybrid Automatic Repeat Request (HARQ) retransmissions. %The \textit{UE Cfg}, \textit{Cell Cfg}, \textit{Sched Cfg}, and \textit{LC Cfg} blocks store the UE configuration, cell configuration, scheduler-specific configuration, and logical channel configuration, respectively. 
We implement SmartCon as an extension to the \texttt{LteEnbMac}. The \texttt{LteEnbPhy} interface reports the SINR of the channel to \texttt{LteEnbMac}, where the SINR is utilized by SmartCon. It uses the subframe block, \texttt{LteAmc}, and \texttt{LteFfrAlgorithm} interfaces to assign subframes with PRBs, MCS, and repetition numbers. %After that, SmartCon utilizes the scheduler interface to schedule packets for transmission.

\subsection{Implementation of Repetitions}

%In order to implement the repetitions feature, 
%the major modifications are carried out in the time domain. %(i.e., the repetition of a data packet). 
Whenever repetition is applied, the successive repetitions of the packets are aggregated at the eNB. We have modified the functionality of the physical layer to incorporate the aggregation of all the repetitions.

%In order to implement repetitions, major modifications are made in the time domain (repeatinga data packet). Whenever repetition is used, the subsequentrepetitions of the same data areaggregated at the eNodeB. Hence, the resulting SNR after theaggregation is the sum of the SNRs of each received repetition. Therefore, repetition oftwo results in an improvement ofapproximately 3dB in SNR [14]. In order to achieve this behavior, we have modified the physicallayer of the base station in NS-3 to aggregate all the repetitions, and use the final sum of SNRas input to the error model described in [15].

\subsection{Training Environment Setup}

%To ensure that the prediction model is not biased by a particular setup, we use a different setup for the training and evaluation of the model. 
%For the training setup, the network has three eNBs in three hexagonal cells with a cell radius of $1.5$ km. Multiple UEs are attached to each eNB. 
For the training setup, the network has three eNBs and each cell has $50$ UEs randomly located inside a cell. %We use {\em proportional fair scheduling} to schedule packets. 
We randomly choose the run time for each simulation instance between $100$-$500$ seconds. The number for runs of each simulation instance is also selected randomly between $60$-$120$, and both downlink and uplink transmissions are considered in every simulation instance. The SINR value is selected randomly between $20$dB-$45$dB and the per-frame SINR is reported by the \texttt{LteEnbMac} interface. The collected data includes the time-stamped packet scheduling (uplink and downlink) events, SINR of the channel, the selected MCS and repetition number, the number of PRBs used for the transmission, and the packet loss rate. The total size of the dataset is close to $2$ GB. %We use this dataset to develop the pre-trained model for SmartCon by following a training procedure discussed in what follows.

\subsection{Training of the Model}
\label{training}

%We need a dataset containing the scheduling events associated with the dynamic selection of the best possible MCS and repetition number considering the present SINR value of the channel. Then we use this dataset to train the proposed GAN model such that the GAN can generate the real dynamics of the probability of packet scheduling ($\alpha_k^{*}(t_l)$), the required number of PRBs ($\gamma_k^{*}(t_l)$) and the best suited MCS and repetition number ($\delta_k^{*}(t_l)$). 
%To generate the dataset for the training purpose, we have applied our MAB model (Algorithm~\ref{algo1}) discussed in Section~\ref{sec:MAB}. 
To generate the data for the dataset, we use the following information -- (i) uplink and downlink scheduling time-stamps, (ii) SINR of the channel, (iii) MCS and repetition number selected for data transmission, (iv) number of PRBs used for the transmission, and (v) average packet loss rate after the transmission. All this data is represented as a time series. 
The training dataset contains information related to the scheduling events, and the selection of the PRBs, MCS, and repetitions. In the dataset, the first parameter identifies the scheduling status of a packet. When a packet is scheduled for an uplink or downlink transmission, the status is ON; otherwise, the status is OFF. When the scheduling status is ON, the number of PRBs used to transmit the packets is determined by the second parameter. The third parameter is the pair of MCS and repetition numbers, which are associated with the scheduling of packets, and thus the third parameter is defined when the scheduling status is ON.

In particular, the dataset used to train the GAN contains $\alpha_k^r(t_l)$, $\gamma_k^r(t_l)$, and $\delta_k^r(t_l)$. Since $\gamma_k^r(t_l)$ stores the normalized value of the number of PRBs and $\delta_k^r(t_l)$ represents the normalized values of the MCS and repetition number, we need to normalize the number of PRBs, MCS value, and repetition number between $[0,1]$. For instance, the MCS value is normalized between $[0,1]$ using $(x-MCS_{min})/(MCS_{max}-MCS_{min})$. In this case, $x$ is considered as the MCS value selected at any time instant. Here, $MCS_{min}$ and $MCS_{max}$ denote the minimum and maximum MCS values in NB-IoT, respectively, where $MCS_{min}=0$ and $MCS_{max}=12$. %The dataset for generating the trained model of the GAN is available at \url{http://dx.doi.org/10.21227/q4vd-qg73}.
%Similarly, the PRBs and repetition number are normalized based on their available maximum and minimum values. 
%\textcolor{blue}{All this data is represented as a time series.} 

After generating the training dataset, the GAN is trained, tested, and validated using $60\%$, $20\%$, and $20\%$ of the dataset, respectively. We have applied a $10$-fold cross-validation technique with randomly chosen validation sets to evaluate the predictive model. In case of the convergence of the GAN, the losses of the discriminator and generator become quite stable after approximately $2200$ epochs. 
After evaluating the predictive model, all the eNBs are loaded with the trained model. Then, the MAB model is again run to collect the same aforesaid information which is used to retrain the GAN model. %This retraining helps in fine tuning the training parameters on the basis of run-time statistics as the network performance in a mobile environment is affected by rapid fluctuations of the signal strength under mobility. The final trained model is loaded in the eNB and used for the performance analysis of SmartCon.
%The final trained model is loaded in the eNB.

\subsection{Prediction Performance}

%We have used $20\%$ of the dataset (test data) to calculate the accuracy of the predictive model. 
To calculate the accuracy of the predictions of the GAN model, we measure the Mean Absolute Percentage Error (MAPE)~\cite{saha2019learning}. MAPE is a continual-time metric that computes the mean absolute deviation between the actual values and the predicted values of the number of PRBs, MCS, repetition number, and probability of scheduling, up to the present time-stamp. We compute the average MAPE value ($\text{MAPE}_{avg}$) of these parameters. We have observed that, for the test data, the $\text{MAPE}_{avg}$ of the GAN model is $5.21\%$ which signifies that the error of the trained model is quite low.

\subsection{Model Size Optimization}

%The size of the trained model significantly depends on the size of the training dataset. Therefore, we need to identify the size of the optimal model. For this purpose, 
%We have applied $10$-fold cross validation on the model by varying the size of the training data. 
Table~\ref{tab:valid} summarizes the observations including the model size and MAPE values ($\text{MAPE}_{avg}$ values) for different training data sizes. From this table, it is noted that $60\%$ of the data from the collected dataset provides a $\text{MAPE}_{avg}$ value of $5.83\%$ with a trained model size of $47.3$ MB. This $\text{MAPE}_{avg}$ is quite acceptable and is associated with a low trained model size ($47.3$ MB). Therefore, we select $60\%$ of the data from the collected dataset to choose our optimal model size. This trained model is loaded in the eNB for online prediction during the execution of SmartCon.
 
\begin{table}[!ht]
    \caption{Model size and $\text{MAPE}_{avg}$ for different training data sizes}
    \label{tab:valid}
    \scalebox{0.88}{
    \begin{tabular}{|l|l|l|l|>{\columncolor[gray]{0.8}}l|l|l|}
        \hline 
        \textbf{Training data size} & $30\%$ & $40\%$ & $50\%$ & \textbf{$60\%$} & $70\%$ & $80\%$ \\
        \hline 
        \textbf{$\text{MAPE}_{avg}$} & $23.8\%$ & $11.4\%$ & $8.56\%$ & \textbf{$5.83\%$} & $5.34\%$ & $4.97\%$\\
        \hline 
        \textbf{Model size (MB)} & $36.2$ & $41.6$ &  $44.8$ & \textbf{$47.3$} & $76.1$ & $129.5$ \\
        \hline 
    \end{tabular}}
\vspace{-5mm}
\end{table}

\subsection{Model Retraining}

%We need to retrain the model such that it can cope with new environments which have not been assimilated during the training of the model. For this purpose, 
%During the run-time of SmartCon, the eNB prepares the dataset by considering the information mentioned in Section~\ref{training}. %In SmartCon, since the selection of the MCS, repetitions, and number of PRBs is based on the minimization of the packet loss rate, the primary idea of retraining the model concentrates on the packet loss rate. 
Whenever a packet loss occurs, SmartCon finds out the correlation between the average packet loss rates in the past execution of duration (window) $\rho$ and the pre-loaded training sample chosen randomly. If this correlation is low, SmartCon sends a signal to the eNB, which signifies that a new sample dataset has been prepared over a window $\rho$. %When a good volume of the sample dataset is geneated, the GAN model is retrained and the eNB is updated with the newly trained model. 
In the implementation, $\rho$ is set to $1$ minute, and the new dataset's size should be of $2$ GB in order to update the trained model. %in the eNB. 

%mechanism discussed in~\cite{zhang2019active}. 

%To explore a new environment that has not been incorporated during pre-training, we use an active semi-supervised learning mechanism discussed in~\cite{zhang2019active} to retrain the DCPP model whenever required. The core idea is as follows. Whenever a frame error occurs, SmartBond finds out an inter-instance correlation between the communication environment (SNR, frames and acknowledgments overheard from neighboring APs with their used CB levels and frame error rates) of past $\Delta$ window and a pre-loaded randomly chosen training sample. If this correlation is low, then the router sends a feedback to the training server with the sample data for this $\Delta$ window. We collect these samples, and when a good amount of samples are received, DCPP is retrained and the routers are updated with the new trained model. In our implementation, we consider $\Delta$ as $1$ minute, and the pre-loaded training sample is of size $1$MB that contains for around $1200$ hours of training data, meticulously chosen containing the instances of different scenarios as captured through an auto-correlation mechanism discussed in~\cite{zhang2019active}. 

\subsection{The Core Module}

The core functionality of SmartCon is to emulate the GAN model on the basis of Ogata's thinning algorithm~\cite{ogata1981lewis}. %If the proposed model is executed continuously, the system load would increase. Therefore, to reduce this load, 
We run SmartCon once in each window $\rho$. %, where the model computes the probabilities of scheduling, the number of PRBs, the MCS values, and the number of repetitions required for the next window $\rho$.

%The core operation of SmartBond is a simple emulation of DCPP based on Ogata's thinning algorithm~\cite{ogata1981lewis}. The system initiates as the standard IEEE 802.11 DBO. To reduce the system load, we do not run SmartBond core for each frame, rather execute it once in every window of size $\Delta$, where it computes the channel bonding levels for the next $\Delta$ window. For every computation window, the system takes the driver (\texttt{ath10k}) reported SNR, FER during previous $\Delta$ window along with the time-stamped information of the CB levels used by neighboring nodes (as per the available knowledge from virtual carrier sensing), and feeds this information as the noise prior to the generator. The generated CB levels are used for the next $\Delta$ window. 

\section{Performance Analysis}
\label{sec:performance}

%To analyze the performance of SmartCon, 
We run each simulation instance for $500$ seconds, where the results are shown as an average of $100$ runs of each simulation instance. Every simulation instance is a combination of downlink and uplink transmissions. %Details of the impact of SmartCon on the performance of NB-IoT are presented next.

\subsection{Analysis of Throughput}

\begin{figure}[!t]
 \centering
 \includegraphics[width=\linewidth]{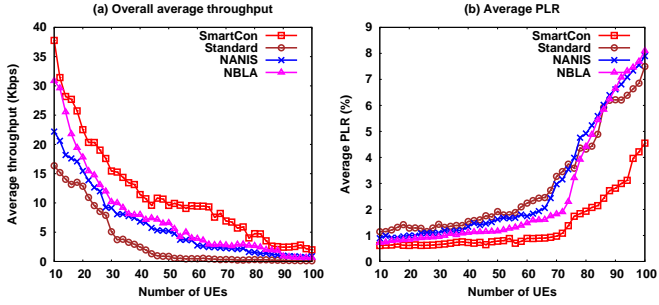}
 \caption{(a) Average throughput and (b) Average packet loss rate}
 \label{fig:g1}
 \vspace{-5mm}
\end{figure}

%In this subsection, we compute the MAC layer throughput at the eNB. Fig.~\ref{fig:g1}(a) shows the result of the average throughput, where it can be observed that SmartCon has a significantly higher throughput than other baseline mechanisms. 
%In SmartCon, the online learning mechanism (MAB) generates a dataset containing the dynamic selection of PRB, MCS levels, and number of repetitions for the transmission of packets during the uplink and downlink scheduling based on the present SINR of the channel. Since this dataset is used to train the propsoed GAN model, 
In SmartCon, the generated dynamics lead to the adaptive selection of PRB, MCS, and repetitions in future scheduling. This learning-based adaptability helps tune the aforesaid parameters based on the present channel condition, such that the average throughput is significantly enhanced. For instance, a higher number of repetitions is chosen when the channel condition is poor so that the transmitted data can reach the destination. %Conversely, it would not be beneficial to apply repetitions when the UE has good coverage. 
Since NBLA is based on threshold-based scheme to adapt the MCS and repetitions, the adaptation is not as efficient as our proposed online learning mechanism. Whereas, since NANIS and the standard approaches do not dynamically deal with the selection of MCS and repetitions, values of these parameters cannot be adaptively tuned in different channel conditions. Therefore, the average throughput is significantly lower in the baselines compared to SmartCon. Fig.~\ref{fig:g1}(a) indicates that, SmartCon has approximately $18$, $2.2$, and $3$ times higher average throughputs than the standard, NANIS, and NBLA schemes, respectively.

\subsection{Analysis of Packet Loss Rate}

%SmartCon specially focuses on reducing the packet loss rate. 
In the generated dataset, the MCS, number of PRBs, and repetition number are chosen to minimize the packet loss rate. In this regard, the application of the best possible value of MCS plays a key role, where a low MCS level should be chosen when the channel condition is poor. Otherwise, the packet loss rate increases. Fig.~\ref{fig:g1}(b) shows that SmartCon has a significantly lower PLR than other baseline mechanisms. When the number of UEs is $50$, the average PLR in SmartCon is approximately $59\%$, $52\%$, and $33\%$ lower than the standard, NANIS, and NBLA approaches, respectively. However, as the number of UEs increases in the network, the adaptability of NANIS and NBLA decreases, as illustrated in Fig.~\ref{fig:g1}(b).   

\subsection{Analysis of Packet Delay}

\begin{figure}[!t]
 \centering
 \includegraphics[width=\linewidth]{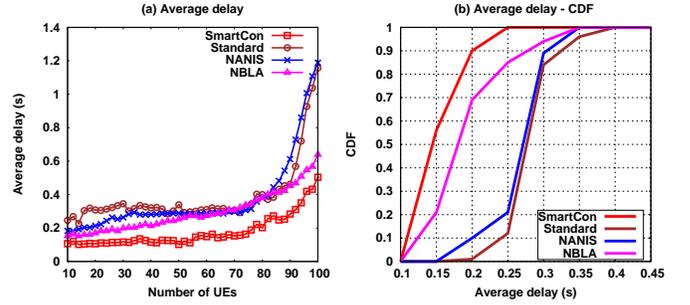}
 \caption{(a) Average packet delay and (b) Average packet delay distribution}
 \label{fig:g2}
 \vspace{-5mm}
\end{figure}

%Packet delay is defined as the average time needed to transmit a packet from the transmitter to the receiver. SmartCon generates the real dynamics of the MCS, number of PRBs, and repetitions selection by studying the pattern of such selection in a wireless environment, such that a packet can be transmitted with the optimized values of these parameters. Particularly, 
In SmartCon, the unnecessary use of repetitions in a transmission helps reduce the time for a packet to reach its destination. Based on the present channel condition, the selection of the best MCS value provides the best possible data rate, and consequently the transmission delay is decreased. From Fig.~\ref{fig:g2}(a), it can be noted that SmartCon has approximately $56\%$, $58\%$, and $21\%$ lower average packet delay than the standard, NANIS, and NBLA schemes, respectively.
Fig.~\ref{fig:g2}(b) illustrates the cumulative distribution function (CDF) of the average packet delay, where the distribution in SmartCon is concentrated in the $0.1-0.25$s range. %Although NBLA adapts the MCS and repetitions, the adaptation is not as efficient as in our proposed online learning mechanism. Thus, 
NBLA provides a higher distribution of average delay ($0.1-0.35$s) than SmartCon, as shown in Fig.~\ref{fig:g2}(b). Whereas, the other baselines have significantly higher average delay CDF (up to $0.4$s). 

\subsection{Analysis of the Number of Consumed Subframes}

\begin{figure}[!t]
 \centering
 \includegraphics[width=\linewidth]{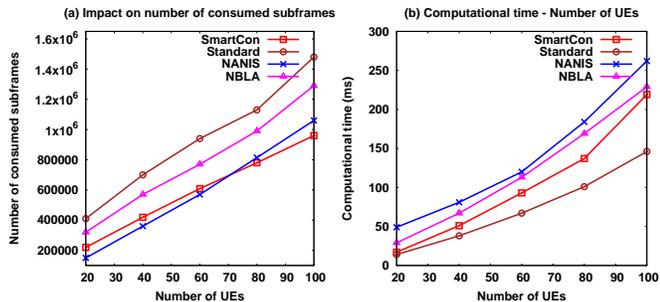}
 \caption{(a) Number of consumed subframes and (b) Computational time}
 \label{fig:g3}
 %\vspace{-5mm}
\end{figure}

%The number of subframes consumed for the transmission is influenced by the number of PRBs selected during the scheduling. 
%SmartCon adjusts the number of PRBs by considering the network condition, and therefore it has a higher dynamic utilization of radio resources. 
%In addition, the tendency of reducing the packet loss rate helps decrease the loss of resources in SmartCon. However, in threshold-based approaches, other competing mechanisms cannot optimize the use of PRBs as performed by SmartCon. 
In the proposed mechanism, the learning is based on the training with a large number of UEs, where the number of sufficient subframes is dynamically adjusted to minimize the PLR. Therefore, based on the training, SmartCon becomes intelligent to appropriately select the number of subframes in different network scenarios having different number of UEs.
%which leads to the adaptation with a less number of subframes in a congested network scenario. 
As a result, from Fig.~\ref{fig:g3}(a), it can be observed that SmartCon has a consumption of subframes approximately $35\%$, $9\%$, and $26\%$ lower than the standard, NANIS, and NBLA schemes, respectively.

\subsection{Analysis of Computational Time}

%As the number of UEs increases in the network, more UEs need to be selected during the scheduling, which results in an increase of the computational time of the scheduling algorithm. 
%From Fig.~\ref{fig:g3}(b), it can be observed that the simple FIFO-based scheme has the lowest computational time. However, 
%The execution time of NANIS is strongly dependent on the count of NPDSCH subframes and the number of UEs considered during the scheduling, where a congested network demands a higher number of NPDSCH subframes. 
In SmartCon, after the training phase, the GAN model simply generates the future dynamics for a UE in the execution phase, which does not depend on any database or set of computations as required in NANIS and NBLA. Therefore, SmartCon has a lower computational time (Fig.~\ref{fig:g3}(b)). When the number of UEs is $100$, SmartCon requires approximately $11\%$, and $4\%$ lower computational time than NANIS, and NBLA, respectively. Meanwhile, when the number of UEs is $50$, in SmartCon, the computational time is approximately $16\%$ lower than NANIS. % when the number of UEs is $50$.

%Next, we analyze the run-time performance of DCPP from our test-bed. Using the data collected from the test-bed during the first seven days of the experiment, we observed that only for $~3.26\%$ of scenarios on average, DCPP made a wrong prediction about the CB levels of the neighboring nodes of an AP. Once we have done a retraining of the model after seven days using the active learning strategy, this error value dropped to $1.68\%$ for the remaining seven days. Further, we observed that SmartBond, on average, takes $10\%-28\%$ of the AP's CPU time and $~95$MB of memory (Archer C2600 has $512$MB of memory) on average. 
%In a nutshell, we observe that DCPP can improve the channel access performance by a large margin during DBO, while not incurring any significant overhead to the router. 

\subsection{Analysis of Selection of MCS Levels}

\begin{figure}[!t]
 \centering
 \includegraphics[width=\linewidth]{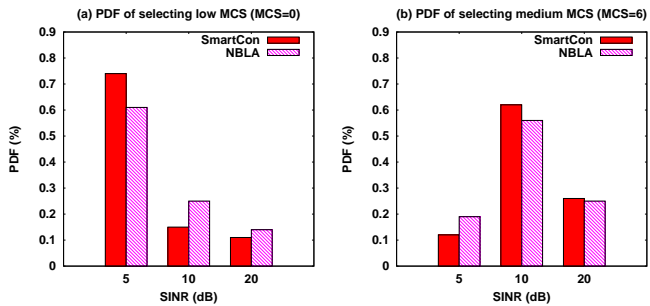}
 \caption{Distribution of the selection of: (a) Low MCS (MCS=$0$) and (b) Medium MCS (MCS=$6$)}
 \label{fig:g4}
 \vspace{-5mm}
\end{figure}
\begin{figure}[!t]
 \centering
 \includegraphics[width=\linewidth]{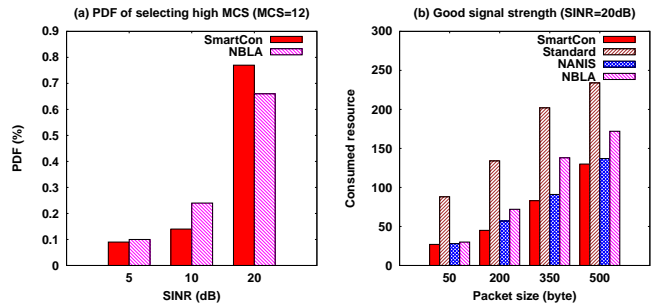}
 \caption{(a) Distribution of the selection of high MCS (MCS=$12$) and (b) Consumed resources under high signal strength (SINR=$20$dB)}
 \label{fig:g5}
 %\vspace{-5mm}
\end{figure}

%Thanks to knowledge on the adaptation of MCS under different network scenarios, the proposed mechanism always tries to dynamically generate the future dynamics of the MCS levels, which results in choosing the best MCS based on the signal quality of the channel. 
Since a higher MCS value increases the PLR in case of low signal strength, the tendency of SmartCon is to decrease the MCS level as the signal strength of the channel deteriorates and vice versa, as illustrated in Figs.~\ref{fig:g4} and~\ref{fig:g5}(a). In our baseline mechanisms, only NBLA adapts the MCS and repetitions, and therefore we consider only NBLA in the analysis of MCS selection.
Table~\ref{table1} presents a comparative analysis of the probability density function (PDF) of the MCS selection in SmartCon and NBLA. %, in low (SINR = $5$ dB), medium (SINR = $6$ dB), and high (SINR = $12$ dB) signal strengths, as shown in Figs.~\ref{fig:g4} and~\ref{fig:g5}(a).   
\begin{table}[h]
\caption{Analysis of the PDF of MCS Selection With Respect To NBLA}
\scriptsize
\centering
\begin{tabular}{|p{1.85cm}|p{1.7cm}|p{1.7cm}|p{1.7cm}|}
\hline
\textbf{MCS Level} & \textbf{SINR = $5$ dB} & \textbf{SINR = $10$ dB} & \textbf{SINR = $20$ dB}\\
\hline
\hline
PDF of MCS=$0$ & $21.31\%$ higher & $40\%$ lower & $21.43\%$ lower\\  
\hline 
PDF of MCS=$6$ & $36.84$ lower & $10.71\%$ higher & $4\%$ lower\\ 
\hline
PDF of MCS=$12$ & $10\%$ lower & $41.67\%$ higher & $16.67\%$ higher\\
\hline
\end{tabular} 
\label{table1}
\vspace{-5mm}
\end{table}

\subsection{Analysis of Consumed Resources under Variable Packet Sizes}

\begin{figure}[!t]
 \centering
 \includegraphics[width=\linewidth]{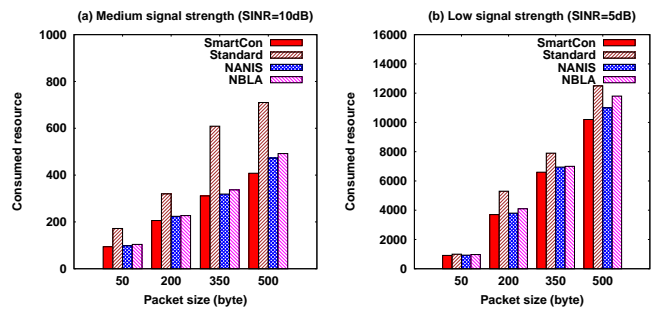}
 \caption{(a) Consumed resources under medium signal strength (SINR=$10$dB) and (b) Consumed resources under low signal strength (SINR=$5$dB)}
 \label{fig:g6}
 \vspace{-5mm}
\end{figure}

%The use of more repetitions enhances the reliability of the transmission at the cost of a spectral efficiency loss. Thus, it is crucial to appropriately use radio resources integrated with the best possible selection of the number of repetitions in NB-IoT. In this regard, SmartCon outperforms the baseline schemes since it 
%SmartCon simultaneously adapts both the PRBs and repetitions along with the MCS level. 
In SmartCon, since the adaptation is performed based on an intelligent prediction considering the present channel condition, the best possible number of resources are chosen dynamically. %and used for scheduling a set of packets. %Therefore, SmartCon makes better use of the radio resources than other baseline mechanisms, as shown in Figs.~\ref{fig:g5}(b) and~\ref{fig:g6}, where we consider three types of signal strengths -- high, medium, and low. 
From Figs.~\ref{fig:g5}(b) and~\ref{fig:g6}, it is noted that, when the packet size is $500$ bytes for an SINR of $20$ dB, SmartCon consumes approximately $1.8$, $1.11$, and $1.32$ times less resources than the standard, NANIS, and NBLA mechanisms, respectively. In case of low signal strength (SINR=$5$dB), the resource consumption in SmartCon are $1.23$, $1.12$, and $1.16$ times lower than the standard, NANIS, and NBLA, respectively.

\subsection{Selection of the MCS and Repetitions by SmartCon}

\begin{figure}[!t]
 \centering
 \includegraphics[width=\linewidth]{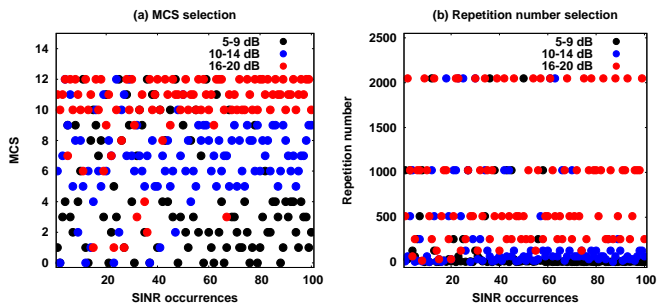}
 \caption{(a) MCS selection and (b) Repetition number selection}
 \label{fig:g7}
 %\vspace{-5mm}
\end{figure}

Fig.~\ref{fig:g7} shows the adaptation of the MCS and repetitions in SmartCon, considering the present channel condition. We capture the values of the MCS and repetition numbers selected for the SINR values of the channel. Since we consider several measurements of the aforementioned parameters against the SINR values, we denote such measurements as `SINR occurrences'. In Fig.~\ref{fig:g7}, the results are shown in three SINR buckets to demonstrate the impacts of the low, medium, and high signal strength. From Fig.~\ref{fig:g7}(a), it is noted that, higher MCS values are selected for the high SINR values, whereas the MCS level decreases as the signal strength deteriorates. Similarly, to efficiently use the repetition mechanism, the repetition number needs to be increased as the SINR of the channel increases, as illustrated in Fig.~\ref{fig:g7}(b), where it is noted that lower repetition numbers are chosen when the SINR values decrease.

\subsection{Performance Comparison with GAN-DDQN}

\begin{figure}[!t]
 \centering
 \includegraphics[width=\linewidth]{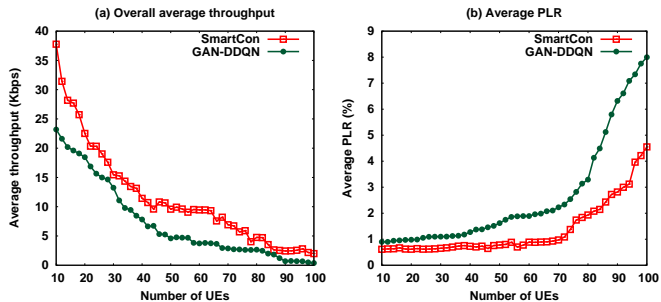}
 \caption{(a) Average throughput and (b) Average PLR}
 \label{fig:g8}
 \vspace{-5mm}
\end{figure}

Fig.~\ref{fig:g8} shows the performance improvement of SmartCon over GAN-DDQN considering the average throughput and packet loss rate (PLR). In particular, GAN-DDQN performs dynamic allocation of radio resources considering network slicing in 5G networks. However, SmartCon intelligently selects the MCS values and repetitions, along with the dynamic adaptation of radio resources. Therefore, in SmartCon, the suitable data rate can be set according to the present channel condition, and consequently the average throughput is improved in SmartCon. Fig.~\ref{fig:g8}(a) shows that SmartCon has approximately an average throughput $5.3$ times higher than the GAN-DDQN. The GAN-DDQN does not specifically handle the reduction of the packet loss in the network, whereas the proposed GAN is trained with a dataset that is intelligently generated by minimizing the average PLR. As a result, SmartCon provides a significantly lower average PLR than the GAN-DDQN. For instance, when the number of UEs is $50$, the average PLR in SmartCon is approximately $52\%$ lower than in GAN-DDQN, as shown in Fig.~\ref{fig:g8}(b).

\section{Conclusion}
\label{sec:concl}

%Based on the point process model, the deep probabilistic learning of SmartCon achieves auto configuration of MCS levels, repetitions, and radio resources in NB-IoT networks, such that the packet loss rate is minimized. 
%Based on the present channel condition, 
The proposed GAN models the stochastic time-stamps of traffic scheduling associated with adaptive MCS values, repetitions, and number of PRBs.
%Specifically, the past history of scheduling events leads to the probability of defining the sequence of time instances of future scheduling associated with the best possible values of the aforementioned parameters, based on the present channel condition. Thus, the training of the GAN-based model is crucial. 
To generate the training dataset for the GAN, we use a MAB-based reinforcement learning mechanism to adapt the MCS, repetitions, and radio resources by considering the present channel condition. %Therefore, the GAN model is trained with an online learning based dataset that leads to smart adaptation in NB-IoT networks. 
%Based on a prototype of SmartCon, 
The detailed simulation analysis demonstrates that SmartCon significantly boosts the performance of NB-IoT networks. The possible limitation of SmartCon is that periodic re-training is required for adjustments under changing network conditions, which led us to apply an active learning approach. 
However, SmartCon provides an important step towards the use of deep generative architecture for the optimization of 5G and B5G networks. 

The future direction of this work can be an intelligent adaptation of the NPDCCH period length along with dynamic adaptation of the MCS, repetitions, and PRBs. The NPDCCH period is defined as the time interval between two successive NPDCCH, where the eNB should allocate the radio resources for the UEs to receive data. The NPDCCH period significantly affects the utilization of the radio resources in NB-IoT networks, and therefore it is required to smartly handle the NPDCCH period when we dynamically adapt the MCS, repetitions, and PRBs.

%SmartBond is based on a deep probabilistic learning over point process model that learns the evolution of channel width selection by the wireless APs. %process from the past history of channel access by the wireless APs. 
%The implementation and thorough testing of SmartBond indicate that it can significantly boost up the channel access performance during IEEE 802.11ac DBO. The possible limitation of the proposed approach is that it requires periodic re-training for the best performance, which we have approached with an active learning methodology. However, active learning may not scale well in real deployments as it require involvement of the vendor for collecting the statistics from the router and retraining the DCPP model; the router further needs an update to replace the old DCPP model with the newly trained one. Nevertheless, we believe that this architecture can work as an important step towards using a deep generative model for improving the online performance of a communication system. 

\section{Acknowledgement}
This work was supported by the Canada Research Chair Program tier-II entitled ``Towards a Novel and Intelligent Framework for the Next Generations of IoT Networks''.

% 
% ----------------------------------------Appendix ------------------------------------------

%\appendices

%\section{}

%-------------------------------------Acknowledgement-----------------------------------

%\ifCLASSOPTIONcompsoc
 % \section*{Acknowledgment*}
%\else
 % \section*{Acknowledgment}
%\fi


\ifCLASSOPTIONcaptionsoff
  \newpage
\fi


%------------------------------------ References -------------------------------------------

\bibliographystyle{IEEEtran}
\bibliography{reference}

%---------------------------------------------------------------------------------------------

%\begin{IEEEbiography[{abc}

%\end{IEEEbiography}


%\newpage



\vfill



\end{document}